\documentclass[review]{elsarticle}

\usepackage{lineno,hyperref}

\usepackage{amsfonts}
\usepackage{color}
\usepackage{bm}

\usepackage{epsfig}
\usepackage{epstopdf}
\usepackage{graphicx,subfigure}

\usepackage{amsfonts}
\usepackage{epsfig}
\usepackage{amssymb}
\usepackage{amsmath}
\usepackage{amsthm}
\usepackage{mathptmx}

\biboptions{authoryear}

\begin{document}

\begin{frontmatter}


\title{Unified nonlocal rational continuum models developed from discrete atomistic equations}

\author[label1]{Amit K. Patra}
\ead{ssamitkpatra@aero.iisc.ernet.in}
\author[label1]{S. Gopalakrishnan}
\cortext[cor1]{Corresponding author.}
\ead{krishnan@aero.iisc.ernet.in}
\author{Ranjan Ganguli\corref{cor1}\fnref{label1}}
\ead{ganguli@aero.iisc.ernet.in}

\address[label1]{Department of Aerospace Engineering, Indian Institute of Science, Bangalore 560012, India}


\begin{abstract}
In this paper, a unified nonlocal rational continuum enrichment technique is presented for improving the dispersive characteristics of some well known classical continuum equations on the basis of atomistic dispersion relations. This type of enrichment can be useful in a wide range of mechanical problems such as localization of strain and damage in many quasibrittle structures, size effects in microscale elastoplasticity, and multiscale modeling of materials. A novel technique of transforming a discrete differential expression into an exact equivalent rational continuum derivative form is developed considering the Taylor's series transformation of the continuous field variables and traveling wave type of solutions for both the discrete and continuum field variables. An exact equivalent continuum rod representation of the 1D harmonic lattice with the non-nearest neighbor interactions is developed considering the lattice details. Using similar enrichment technique in the variational framework, other useful higher-order equations, namely nonlocal rational Mindlin-Herrmann rod and nonlocal rational Timoshenko beam equations, are developed to explore their nonlocal properties in general. Some analytical and numerical studies on the high frequency dynamic behavior of these novel nonlocal rational continuum models are presented with their comparison with the atomistic solutions for the respective physical systems. These enriched rational continuum equations have crucial use in studying high-frequency dynamics of many nano-electro-mechanical sensors and devices, dynamics of phononic metamaterials, and wave propagation in composite structures.
\end{abstract}


\end{frontmatter}


\section{Introduction}
Experimental observations at micro/nanoscales have shown the inability of classical continuum mechanics in providing crucial insight into the basic phenomena of deformation, crack growth, and phase transformation in numerous material systems. Therefore, there is a need for convenient and generalized continuum models which can elucidate the detailed mechanisms of various micro/nano-electromechanical sensors and devices widely used in nano-technology, material science, and biology. Moreover, an efficient continuum equation can very accurately approximate the mean part of atomistic solutions and thus, can substantially improve the accuracy and economy of a multiscale analysis. However, a potent continuum equation should be able to describe certain mechanical phenomena which are due to the intrinsic characteristic of microstructural details of a material system. Unfortunately, as a coarse scale model, the conventional continuum theory fails to do so. In reality, however, no material is an ideal continuum but is discrete in nature with complicated spatially varying internal microstructure. Therefore, the conventional continuum description is no longer adequate to describe mechanical details of any material at these small scales. Thus, enrichment of conventional continuum theories is required to improve the agreement between the continuum predictions and the experimental observations.

Any existent microstructural heterogeneity of crystalline materials affects the dispersion of elastic waves, which is a real physical phenomenon that can be observed and studied experimentally. The shape of a general propagating wave changes over time due to the different propagating speeds of different harmonic components in it. This phenomenon is commonly known as dispersion. In the dynamics context, however, the most important motivation is to improve the dispersive characteristics of continuum equations as compared to the actual atomistic dispersion phenomena. In general, the dispersive characteristic of a material system is mainly dominated by the discrete characteristics of their crystal lattices and the range of interatomic interaction forces prevailing in the system \citep{brillouin2003wave,kittel2008sspbook,jirasek2004nonlocal}. However, these vital dispersion properties of short elastic waves in actual discrete atomistic systems are lost due to the standard homogenization procedure involved in the development of the continuum models \citep{jirasek2004nonlocal}.

Over the last century, many researchers have expended their valuable efforts in constructing several enrichments of the standard continuum theory to capture the important features of the microstructure of any heterogeneous material system. All these variety of enrichment models are classified under generalized nonlocal continuum models. Most of these enrichment techniques are inspired by the seminal idea of the Cosserat brothers \citep{cosserat1909theorie}. All these generalized Cosserat theories consider additional fields which are independent of displacement field and provide supplementary information on the small scale kinematics. One such major class of enrichments is based on the idea that the total elastic energy density is a function of higher order gradients ($u^{''}$, $u^{'''}$, $u^{''''}$, etc.) as well as the first gradient of the displacement fields ($u$) \citep{toupin1962elastic,toupin1964theories,mindlin1962effects,mindlin1964micro,mindlin1968first,kolter1964couple,koiter1969couple}. This class of enrichment is known as strain-gradient theory of elasticity in the field of nonlocal continuum mechanics. The simplest enrichment of this class assumes the strain energy density as the function of $u^{'}$ and $u^{''}$. This results in a wave equation, which has an extra fourth order partial derivative term as compared to the classical wave equation. The dispersion relation of this enriched wave equation, for the positive values of higher-order elastic modulus, becomes erroneously opposite to that of the discrete atomistic model and the experimental observations at higher frequency \citep{jirasek2004nonlocal}. Therefore, this strain-gradient elasticity model can give a reasonable approximation of the actual dispersion characteristics only for some negative values of higher-order elastic modulus. However, the assumption of negativeness of the higher-order elastic modulus gives physically absurd results like the loss of convexity of elastic potential and unstable behavior of short elastic waves. Therefore, the enriched wave equation obtained by this model is termed as the ``bad Boussinesq problem".

Another class of nonlocal enrichments is known as models with mixed spatial-temporal derivatives which ultimately tries to overcome the snags of the ``bad Boussinesq problem". \cite{fish2002non} proposed a method in which they replace the fourth order partial derivative term $u^{''''}$ by a term with mixed derivative, $\ddot{u}^{''}$. This model gives reasonable accuracy in dispersion prediction at moderate frequency and overcomes the problem of instability for very short waves. This is known as the ``good Boussinesq problem" and its extension to multiple dimensions is complicated \citep{fish2002non2}. Metrikine and Askes \citep{metrikine2002onePart1,askes2002onePart2,askes2005higher} proposed a different continualization method of this class which includes an additional heuristic model parameter. With a proper choice of parameter, this model can reasonably approximate dispersion and can avoid instability for short elastic waves. However, this model has a fourth order derivative term $u^{''''}$ and higher-order boundary conditions are required to solve the problem.

The popular integral type nonlocal elasticity is based on weighted spatial averaging. This is also an elegant generalization of Cosserat theories \citep{cosserat1909theorie}. Eringen and Edelen \citep{edelen1971nonlocal,eringen1965linear,eringen1972linear,eringen1983differential} have, mainly, advanced the theories of integral type nonlocal elasticity. In this type of enrichments, the constitutive relation at a point of continuum involves weighted averages of the state variables over a certain neighborhood of that point \citep{bazant2002nonlocal}. In their constitutive relation, they use a dimensionless even function called nonlocal weight function which determines the accuracy of the model. \cite{eringen1983differential} derived the approximate expression of this nonlocal weight function by studying the relationship between 1D atomic lattice with nearest-neighbor interaction and the nonlocal integral models. This has inspired a plethora of research articles on the extension of this model for higher dimensions by several other researchers \citep{peddieson2003application,wang2005waveN,wang2006buckling,reddy2007nonlocal,reddy2008nonlocal,aydogdu2009general,reddy2010nonlocalbeamsplates,
narendar2009nonlocal,narendar2010nonlocal,narendar2010terahertz,narendar2010ultrasonic,narendar2011prediction}.
This model has good reasonable accuracy in predicting the dispersion relation in 1D. However, this too is a reasonably approximate method, but not an exact method. Eringen's integral model would reproduce the dispersion relation for $1$D atomistic model only for wavelengths larger than $2l_a$, where $l_a$ is the lattice spacing \citep{jirasek2004nonlocal}. The expression of weight function, proposed by Eringen, is very complicated and not very practical to use \citep{jirasek2004nonlocal}.


Observing the difficulties in determining microstructure-dependent length scale parameters \citep{lam2003experiments,maranganti2007novel}, \cite{park2006bernoulli,park2008variational} used the modified couple stress theory proposed by \cite{yang2002couple} to develop nonlocal Euler-Bernoulli beam which has only one material length scale parameter. \cite{ma2008microstructure} developed a microstructure-dependent Timoshenko beam model based on the same modified couple stress theory. Afterwards, \cite{reddy2011microstructure} developed microstructure-dependent nonlinear Euler-Bernoulli and Timoshenko beam theories for functionally graded beams using the principle of virtual displacements, which has motivated several applications.

In the current paper, we propose a novel and elegant technique of transforming a discrete differential expression into its exactly equivalent rational continuum derivative form. Using the concept of Taylor's series expansion of sufficiently smooth displacement field $u$ we show that, for a traveling wave type of solution, a discrete mechanical model can exactly be realized as a partial derivative term premultiplied by a linear differential operator with constant coefficients which incorporates the discrete microstructural information. Using this technique, we have derived the exact equivalent wave equation of a 1D harmonic lattice with nearest neighbor interaction as well as with non-nearest neighbor interactions. Most interestingly, the gradient type nonlocal elasticity models \citep{toupin1962elastic,toupin1964theories,mindlin1962effects,mindlin1964micro,mindlin1968first} and the integral type nonlocal elasticity models \citep{edelen1971nonlocal,eringen1965linear,eringen1972linear,eringen1983differential} are shown to be the special cases of this generalized nonlocal rational continuum formulation. Development of some other higher order continuum equations namely nonlocal rational Mindlin-Herrmann rod and nonlocal rational Timoshenko beam equations are shown to be straightforward using this novel idea. These nonlocal rational continuum models have only one microstructure-dependent length scale parameter. However, these equations show substantially improved agreement in the dispersion characteristics with the linearized atomistic equations. Most importantly, these equations are simple and can be solved analytically and/or semi-analytically in the Laplace transform based spectral finite element method (NLSFEM) \citep{murthy2011signal,patra2014spectral} framework with ease. Moreover, simple approximations in the nonlocal rational continuum formulations generate some well known gradient type and/or integral type nonlocal elasticity equations which can be solved analytically and/or using the conventional finite element method.

The rest of this paper is organized as follows. The novel nonlocal enrichment of continuum equations on the basis of atomistic dispersion relation is presented in Section $2$. In Section $2.1$, an exact equivalent nonlocal rational rod realization of 1D harmonic lattice with nearest neighbor interaction as well as with non-nearest neighbor interactions and the derivation of several discrete dispersion differential operators are described in detail in Section $2.1.1$. The variational formalism for the general derivation of nonlocal rational continuum equations with appropriate boundary conditions is presented in Section $2.1.2$. Section $2.2$ describes the details of nonlocal rational Mindlin-Herrmann rod approximation of a 2D harmonic lattice. The details of nonlocal rational Timoshenko beam approximation of a 2D harmonic lattice system is presented in Section $2.3$. Section $3$ presents the analytical and numerical studies. Analytical solutions for dispersive characteristics of a 1D harmonic lattice with nearest neighbor interaction as well as with next-nearest neighbor interactions are presented in Section $3.1$. The comparison of numerical solutions by both the molecular dynamics (MD) and NLSFEM is presented in this section. The dispersive characteristics of nonlocal rational Mindlin-Herrmann rod and nonlocal rational Timoshenko beam are compared with their respective atomistic predictions in Section $3.2$ and Section $3.3$, respectively. The small scale effects on the natural frequencies of a simply supported nonlocal rational Timoshenko beam are studied in Section $3.4$. The paper concludes with a summary in Section $4$.

\section{Unified nonlocal rational continuum enrichment on the basis of atomistic dispersion relation}
The dispersion prediction of elastic waves deteriorates when an actual discrete system is modeled as a homogeneous continuum. Here we describe a novel technique to incorporate the discrete dispersion characteristics of an actual atomistic system in its equivalent continuum equations. In this section, the procedure of finding proper discrete dispersion differential operators and the variational derivation technique of several unified nonlocal rational continuum equations are described.

\begin{figure}[!h]
\centering
\includegraphics[scale=0.4]{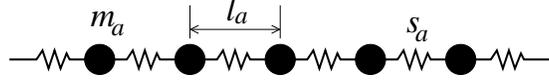}
\caption{$1$D harmonic lattice.}
\label{harmoniclatt}
\end{figure}
\subsection{An equivalent nonlocal rational nod for 1D harmonic lattice: Exact discrete dispersion relation}
 The enriched 1D wave equation is developed directly from the atomistic equation of 1D lattice. For the continuum enrichment, an equivalent 1D harmonic lattice model (Fig. \ref{harmoniclatt}) of any 1D lattice with nonlinear potential is considered. The lattice consists of $n$ identical atoms of equal mass $m_a$ connected by $n$ identical, massless, lossless, perfectly linear springs of stiffness $s_a$. The initial position $x_{j0}$ and deformed position $x_{j}$ of the $j^{th}$ atom are given by
\begin{equation}\label{SERE501}
x_{j0}=jl_a; \;\; \;\; \;\; \;\; x_{j}=jl_a + u_j; \;\; \;\; j=0,1,2,3...,n
\end{equation}
where, $l_a$ is the lattice constant and $u_j$ is the out-of-equilibrium displacement. Force exerted on the $j^{th}$ atom by the $(j+p)^{th}$ atom in the harmonic lattice is proportional to the difference $u_{j+p}-u_j$. First, the formalism is developed considering nearest-neighbor interactions only. Afterwards, the formalism is extended for non-nearest-neighbor interactions. The homogeneous equation of motion of the $j^{th}$ atom ($1<j<n$) in the harmonic lattice obeys Newton's second law as
\begin{equation}\label{SERE502}
m_a\ddot{u}_j = s_a(u_{j+1} - 2u_j + u_{j-1})
\end{equation}
Considering the traveling wave solution of the type $u_j=A e^{i(k_ljl_a-\omega t)}$, the dispersion relation and the expression of group velocity $C_g$, and phase velocity $C_p$ are given by \citep{kittel2008sspbook}
\begin{align}\label{SERE503}
 \omega^2= 2\frac{s_a}{m_a}(1 - \cos k_ll_a)\nonumber\\
 C_p = \frac{\omega}{k_l} = l_a\sqrt{\frac{s_a}{m_a}} \left|\frac{\sin \frac{k_ll_a}{2}}{\frac{k_ll_a}{2}}\right|\nonumber\\
 C_g = \frac{d\omega}{dk_l} = l_a\sqrt{\frac{s_a}{m_a}} \left| \cos \frac{k_ll_a}{2}\right|
 \end{align}
where $k_l$ and $\omega$ are the longitudinal wavenumber and the circular frequency, respectively and $\left|.\right|$ denotes absolute values. Now assuming the 1D chain of atoms as a continuum rod of uniform cross-section $A$ with Young's modulus $E$ and density $\rho$, letting $m_a=\rho A l_a$ and $s_a = EA/l_a$, traveling wave solution will be of the type $u(x,t)=A e^{i(k_lx-\omega t)}$. Here, the discrete solution $u_j = A e^{i(k_ljl_a-\omega t)} = A e^{i(k_lx_j-\omega t)} = u(x_j,t)$ is a subset of the continuum solution $u(x,t)$. Considering this, one can rewrite equation (\ref{SERE502}) as
\begin{equation}\label{SERE5102}
\frac{\partial^2 u_j}{\partial t^2} = \frac{E}{\rho} \frac{(u_{j+1} - 2u_j + u_{j-1})}{{l_a}^2}
\end{equation}
The equation of motion (\ref{SERE5102}) can be reduced to the continuum limit equation for the displacement field $u(x,t)$, as described by \cite{patra2014spectral}, considering $l_a\rightarrow 0$ as
\begin{equation}\label{SERE5202}
E\frac{\partial^2 u(x,t)}{\partial x^2} - \rho \frac{\partial^2 u(x,t)}{\partial t^2}=0
\end{equation}
Or, the homogenized Eq. (\ref{SERE5202}) can be expressed as
\begin{equation}\label{SERE5202L}
\textbf{L}u(x,t)=\textbf{f}
\end{equation}
where $\textbf{f}=0$ denotes the right-hand side external forcing and $\textbf{L}$ denotes the left-hand side homogeneous differential operator in Eq. (\ref{SERE5202}). The dispersion relation and the expression of group velocity $C_g$, and phase velocity $C_p$ associated with Eq. (\ref{SERE5202}) are given as
\begin{align}\label{SERE5302}
\omega^2= \frac{E}{\rho} {k_l}^2\nonumber\\
C_p = \frac{\omega}{k_l} = \sqrt{\frac{E}{\rho}} =  l_a\sqrt{\frac{s_a}{m_a}}\nonumber\\
C_g = \frac{d\omega}{dk_l} = \sqrt{\frac{E}{\rho}} =  l_a\sqrt{\frac{s_a}{m_a}}
\end{align}
The deterioration in the above expressions of dispersion relation and group velocity (Eq. (\ref{SERE5302})) from its original atomistic one (Eq. (\ref{SERE503})) is mainly due to the limiting consideration of $l_a\rightarrow 0$. However, the atomic separation $l_a$, which is of the order of $10^{-10}$ m in reality, is many order bigger unit of length as compared to the smallest possible Plank's length $l_p = 1.616199(97) \times 10^{-35}$ m considered in physics. Therefore, the consideration of $l_a\rightarrow 0$ is wrong for a realistic atomistic systems. The limiting equation (\ref{SERE5202}) gives erroneous results for actual 1D harmonic lattices other than the long wave condition.

To overcome this lacuna, the exact equivalent continuum equation is derived in a different manner. Considering the Taylor's series expansion of the sufficiently smooth function $u(x,t)$ and for $\Delta x = l_a$, one can write the discrete differential portion of the right-hand side of Eq. (\ref{SERE502}) as
\begin{equation}\label{SERE504}
\frac{u(x+l_a,t) - 2u(x,t) + u(x-l_a,t)}{{l_a}^2} = \frac{\partial^2 u(x,t)}{\partial x^2} + O({l_a}^2)
\end{equation}
Here, $O({l_a}^2)$ is an infinite sum and cannot be neglected when $l_a\nrightarrow 0$. Therefore, the participation of $O({l_a}^2)$ is taken into account by premultiplying a discrete dispersion operator $\Delta_m$ with the partial derivative and the Eq. (\ref{SERE504}) is rewritten as
\begin{equation}\label{SERE505}
\frac{\partial^2 u(x,t)}{\partial x^2} + O({l_a}^2) = \Delta_2 \frac{\partial^2 u(x,t)}{\partial x^2} = \frac{u(x+l_a,t) - 2u(x,t) + u(x-l_a,t)}{{l_a}^2}
\end{equation}
Here the subscript $m$ of $\Delta_m$ denotes the order of the partial derivatives of the field variables with which it is pre-multiplied. Considering the continuum traveling wave solution $u(x,t)=A e^{i(k_lx-\omega t)}$, Eq. (\ref{SERE505}) can be reduced to the form
\begin{equation}\label{SERE506}
\Delta_2 \frac{\partial^2 (A e^{i(k_lx-\omega t)})}{\partial x^2} = \left(\frac{(e^{ik_lx} - 2 + e^{-ik_lx})}{{l_a}^2}\right) A e^{i(k_lx-\omega t)} = - 2\frac{(1 - \cos k_ll_a)}{{l_a}^2} A e^{i(k_lx-\omega t)}
\end{equation}
Now manipulating the Eq. (\ref{SERE506}) as in the following manner, the modified equation is rewritten as
\begin{align}\label{SERE507}
\Delta_2 \frac{\partial^2 (A e^{i(k_lx-\omega t)})}{\partial x^2} = - \left(\frac{\sin \frac{k_ll_a}{2}}{{\frac{l_a}{2}}}\right)^2 \frac{\partial^2 }{\partial x^2} \iint (A e^{i(k_lx-\omega t)}) \, dx \, dx \nonumber\\
= \left(\frac{\sin \frac{k_ll_a}{2}}{{\frac{k_ll_a}{2}}}\right)^2 \frac{\partial^2 (A e^{i(k_lx-\omega t)})}{\partial x^2}= \left(1-\frac{{k_l}^2 {l_a}^2}{12}+\frac{{k_l}^4 {l_a}^4}{360}- ....\right)\frac{\partial^2 (A e^{i(k_lx-\omega t)})}{\partial x^2}
\end{align}
The Eq. (\ref{SERE507}) can be reduced as
\begin{equation}\label{SERE507ex}
\Delta_2 \frac{\partial^2 (A e^{i(k_lx-\omega t)})}{\partial x^2} = \left(1-\frac{{l_a}^2}{12{i}^2}\frac{\partial^2}{\partial x^2}+\frac{{l_a}^4}{12{i}^4}\frac{\partial^4}{\partial x^4}- ....\right)\frac{\partial^2 (A e^{i(k_lx-\omega t)})}{\partial x^2}
\end{equation}
From which, it follows that
\begin{equation}\label{SERE50777}
\Delta_2  = \left(1-\frac{{l_a}^2}{12{i}^2}\frac{\partial^2}{\partial x^2}+\frac{{l_a}^4}{12{i}^4}\frac{\partial^4}{\partial x^4}- ....\right)
\end{equation}
The linear differential operator $\Delta_2$ contains infinite terms with constant coefficients. It is seen from Eqs. (\ref{SERE504}) - (\ref{SERE507ex}) that the discrete differential expression in Eq. (\ref{SERE504}) can be transformed into an exact derivative term with a multiplier linear differential operator with constant coefficients without any approximation. This novel transformation is true for any non-zero value of $l_a$.

The solution $u(x,t)$ below the resolution of atomistic solution $u(x_j,t)$ is physically meaningless. Therefore, considering $u(x,t)$ as an equivalent representation of $u(x_j,t)$ and using the development of Eqs. (\ref{SERE504}) - (\ref{SERE50777}), Eq. (\ref{SERE502}) is rewritten as
\begin{align}\label{SERE508}
\frac{\partial^2 (A e^{i(k_lx-\omega t)})}{\partial t^2} = {l_a}^{2}\frac{s_a}{m_a} \left(\frac{\sin \frac{k_ll_a}{2}}{{\frac{k_ll_a}{2}}}\right)^2 \frac{\partial^2 (A e^{i(k_lx-\omega t)})}{\partial x^2} = \frac{E}{\rho} \left(\frac{\sin \frac{k_ll_a}{2}}{{\frac{k_ll_a}{2}}}\right)^2 \frac{\partial^2 (A e^{i(k_lx-\omega t)})}{\partial x^2}
\end{align}
From which, it follows that
\begin{align}\label{SERE508ex}
\frac{\partial^2 u(x,t)}{\partial t^2} = \frac{E}{\rho} \left(1-\frac{{l_a}^2}{12{i}^2}\frac{\partial^2}{\partial x^2}+\frac{{l_a}^4}{12{i}^4}\frac{\partial^4}{\partial x^4}- ....\right) \frac{\partial^2 u(x,t)}{\partial x^2} = \frac{E}{\rho} \Delta_2 \frac{\partial^2 u(x,t)}{\partial x^2}
\end{align}
Or, the nonlocal rational continuum rod Eq. (\ref{SERE508ex}) can be expressed as
\begin{equation}\label{SERE508Lbar}
\bar{\textbf{L}}u(x,t) = \textbf{f}
\end{equation}
where $\bar{\textbf{L}}$ denotes the corresponding nonlocal rational differential operator in Eq. (\ref{SERE508Lbar}) and $\textbf{f}=0$ for no external forcing. This new enriched wave equation (\ref{SERE508ex}) has exactly identical dispersion relation as the atomistic Eq. (\ref{SERE503}) and can exactly represent its atomistic counterpart for all wavenumbers/frequencies. The novel nonlocal rational rod model contains one additional microstructural length scale parameter $l_a$. The enriched continuum equation (\ref{SERE508}) reduces to it's classical form at the long wave limit i.e., when $k_l l_a\rightarrow 0$. This equation is very simple to solve in the frequency domain and can be solved semi-analytically using NLSFEM to give very accurate predictions about discrete atomistic systems.

For small frequency/long wave length condition (i.e., when $k_a l_a$ is small), neglecting higher degree terms above the term containing ${k_l}^4 {l_a}^4$, Eq. (\ref{SERE507ex}) can be approximated as
\begin{equation}\label{SERE5077exStarinG}
\Delta_2 \frac{\partial^2 (A e^{i(k_lx-\omega t)})}{\partial x^2} \approx \left(1-\frac{{l_a}^2}{12{i}^2}\frac{\partial^2}{\partial x^2}\right)\frac{\partial^2 (A e^{i(k_lx-\omega t)})}{\partial x^2}
\end{equation}
From which, the approximate equivalent version of Eqs. (\ref{SERE502}) and (\ref{SERE508ex}) can be obtained as
\begin{align}\label{SERE5088exStarinG}
\frac{\partial^2 u(x,t)}{\partial t^2} = \frac{E}{\rho} \left(1-\frac{{l_a}^2}{12{i}^2}\frac{\partial^2}{\partial x^2}\right) \frac{\partial^2 u(x,t)}{\partial x^2} = \frac{E}{\rho}\frac{\partial^2 u(x,t)}{\partial x^2}+\frac{{l_a}^2}{12}\frac{E}{\rho}\frac{\partial^4 u(x,t)}{\partial x^4}
\end{align}
This approximate Eq. (\ref{SERE5088exStarinG}) resembles the wave equation of strain-gradient elasticity also known as ``bad Boussinesq problem'' \citep{jirasek2004nonlocal} but with the corrected sign. This shows that the corrected wave equation of strain-gradient elasticity is a small wavenumber approximation of the novel nonlocal rational continuum model.

Most of the brittle materials have short range interactions. Therefore, an atomistic model with nearest-neighbor or next-nearest-neighbor interactions can very satisfactorily describe the system dynamics of brittle materials. In case of metals, the range of interactions observed are of quite long range and have been found between as many as $20^{th}$ nearest neighbors \citep{kittel2008sspbook}. For non-nearest-neighbor interactions, the generalized atomistic dispersion relation can be written as \citep{kittel2008sspbook, jirasek2004nonlocal}
\begin{align}\label{SERE509}
 \omega^2 = \frac{2}{m_a} \sum_{p=1}^{N} s_{ap} (1 - \cos k_l p l_a) = \frac{4}{m_a} \sum_{p=1}^{N} s_{ap} \sin^2 \frac{k_l p l_a}{2}
\end{align}
where $s_{ap}$ is the equivalent harmonic spring constant of $p^{th}$ nearest neighbor interaction. For any nonlinear interatomic potential as considered in \cite{rafii2004interatomic}, magnitude of $s_{ap}$ decreases with the increasing values of $p$. For example, in case of a Lennard-Jones type potential, the magnitude of $s_{a2}$ is approximately $2$ order less with respect to the magnitude of $s_{a1}$. Therefore, by straightforward extension, the generalized enriched continuum wave equation for a harmonic lattice with non-nearest-neighbor interactions can be written as
\begin{align}\label{SERE5010}
\frac{\partial^2 (A e^{i(k_lx-\omega t)})}{\partial t^2} = {l_a}^{2}\frac{s_{a1}}{m_a} \sum_{p=1}^{N} p^2 \frac{s_{ap}}{s_{a1}} \left(\frac{\sin \frac{k_lpl_a}{2}}{{\frac{k_lpl_a}{2}}}\right)^2 \frac{\partial^2 (A e^{i(k_lx-\omega t)})}{\partial x^2}
\end{align}
Or, in simplified form, the enriched wave equation is obtained as
\begin{align}\label{SERE5011}
\frac{\partial^2 u(x,t)}{\partial t^2} = {l_a}^{2}\frac{s_{a1}}{m_a} \sum_{p=1}^{N} p^2 \frac{s_{ap}}{s_{a1}}\Delta_{2p} \frac{\partial^2 u(x,t)}{\partial x^2} = \frac{E}{\rho} \sum_{p=1}^{N} p^2 \; R_{p} \; \Delta_{2p} \frac{\partial^2 u(x,t)}{\partial x^2}
\end{align}
where $R_{p} = s_{ap}/s_{a1}$ is the stiffness ratio and $\Delta_{2p} = (1-\frac{{p}^2{l_a}^2}{12{i}^2}\frac{\partial^2}{\partial x^2}+\frac{{p}^4{l_a}^4}{12{i}^4}\frac{\partial^4}{\partial x^4}- ....)$ is the second order discrete dispersion operator for the $p^{th}$ nearest neighbor interaction. Eq. (\ref{SERE5011}) is the exact equivalent continuum representation of the 1D harmonic lattice with $N$ non-nearest neighbor interactions. For the sake of clarity, we rewrite Eq. (\ref{SERE5010}) for the next-nearest neighbor interaction (i.e., for $p=2$) in detail as
\begin{align}\label{SERE51011}
\frac{\partial^2 (A e^{i(k_lx-\omega t)})}{\partial t^2} = \frac{E}{\rho} \left[\left(\frac{\sin \frac{k_ll_a}{2}}{{\frac{k_ll_a}{2}}}\right)^2 +  2^2 R_2 \left(\frac{\sin k_ll_a}{k_ll_a}\right)^2 \right] \frac{\partial^2 (A e^{i(k_lx-\omega t)})}{\partial x^2}
\end{align}
Expanding the $\frac{\sin k_ll_a}{k_ll_a}$ term, one can simplify the Eq. (\ref{SERE51011}) as
\begin{align}\label{SERE52011}
\frac{\partial^2 (A e^{i(k_lx-\omega t)})}{\partial t^2} = \frac{E}{\rho} \left[(1 + 4 R_2) \left(\frac{\sin \frac{k_ll_a}{2}}{{\frac{k_ll_a}{2}}}\right)^2 - 4 R_2 \left(\frac{\sin^2 \frac{k_ll_a}{2}}{{\frac{k_ll_a}{2}}}\right)^2 \right] \frac{\partial^2 (A e^{i(k_lx-\omega t)})}{\partial x^2}
\end{align}
After similar manipulation as in Eq. (\ref{SERE507}), the exact continuum form of the discrete 1D harmonic lattice with next-nearest neighbor interaction is obtained as
\begin{align}\label{SERE53011}
\frac{\partial^2 (A e^{i(k_lx-\omega t)})}{\partial t^2} = \frac{E}{\rho} \left[(1+4 R_2)\left(\frac{\sin \frac{k_ll_a}{2}}{{\frac{k_ll_a}{2}}}\right)^2 + {l_a}^2 R_2 \left(\frac{\sin \frac{k_ll_a}{2}}{{\frac{k_ll_a}{2}}}\right)^4 \frac{\partial^2}{\partial x^2}\right]\frac{\partial^2 (A e^{i(k_lx-\omega t)})}{\partial x^2}
\end{align}
From which, it follows that
\begin{align}\label{SERE53011ex}
\frac{\partial^2 u(x,t)}{\partial t^2} = \frac{E}{\rho} \left[(1+4 R_2)\Delta_{2}\frac{\partial^2 u(x,t)}{\partial x^2} + {l_a}^2 R_2 {\Delta_2}^2 \frac{\partial^4 u(x,t)}{\partial x^4}\right]
\end{align}
Eq. (\ref{SERE53011ex}) clearly shows that the interactions between the non-nearest neighbor atoms are taken into account exactly by the inclusion of higher order derivative terms in the equation. It can be seen that for the limiting condition of $l_a\rightarrow 0$ and for nearest neighbor interaction, Eq. (\ref{SERE53011ex}) reduces to the classical rod equation (\ref{SERE5202}). However, the long wave consideration (i.e., $k_ll_a\rightarrow 0$) reduces the Eq. (\ref{SERE53011}) to a correct form of strain-gradient elasticity equation which resolves the sign paradox of the ``bad Boussinesq problem". Following similar approach as above, exact equivalent continuum wave equations for distant neighbor interactions can be obtained from Eq. (\ref{SERE5011}).

Eq. (\ref{SERE53011ex}), in fact, includes a dependance on the immediate as well as non-nearest neighborhood of the point under consideration. In broad sense, continuum models involving a characteristic length scale parameter or a weighted summation/intregal are generally considered as nonlocal continuum models \citep{bazant2002nonlocal}. These enriched wave equations contain the linear differential operator of infinite terms with constant coefficients, which are functions of the lattice parameter $l_a$. Therefore, these type of enriched continuum formulations (i.e., Eqs. (\ref{SERE508}), (\ref{SERE53011ex})) can be considered as generalized nonlocal continuum models as they contain a characteristic length parameter (lattice constant) and the weighted nonlocal summations. The nonlocal expression proposed by \cite{eringen1983differential} can be shown as an approximation of Eq. (\ref{SERE508ex}).

\subsubsection{Discrete dispersion differential operators}
In the similar fashion, the expressions of some important discrete dispersion differential operators $\Delta_m$s associated with partial derivatives of different orders are derived. The field variable $u(x,t)$ is a very smooth function. Therefore, one can consider the Taylor's series expansion of $u(x,t)$ about any atom locations in the lattice and write
\begin{equation}\label{SERE5012}
\frac{u(x+\frac{l_a}{2},t) - u(x-\frac{l_a}{2},t)}{l_a} = \frac{\partial u(x,t)}{\partial x} + O({l_a}^2) = \Delta_1 \frac{\partial u(x,t)}{\partial x}
\end{equation}
For the same traveling wave solution of the type $u(x,t)=A e^{i(k_lx-\omega t)}$, Eq. (\ref{SERE5012}) reduces to the form
\begin{equation}\label{SERE5013}
\left(i \frac{\sin \frac{k_ll_a}{2}}{\frac{l_a}{2}}\right)  u(x,t) = \Delta_1 \frac{\partial (A e^{i(k_lx-\omega t)})}{\partial x}
\end{equation}
Manipulating the Eq. (\ref{SERE5013}) in the following manner, one can obtain
\begin{equation}\label{SERE5014}
\left(i \frac{\sin \frac{k_ll_a}{2}}{\frac{l_a}{2}}\right) \frac{\partial }{\partial x} \int (A e^{i(k_lx-\omega t)}) \, dx = \left(\frac{\sin \frac{k_ll_a}{2} }{\frac{k_ll_a}{2}}\right) \frac{\partial (A e^{i(k_lx-\omega t)})}{\partial x} = \Delta_1 \frac{\partial (A e^{i(k_lx-\omega t)})}{\partial x}
\end{equation}
Therefore, the first order discrete dispersion differential parameter is obtained as
\begin{equation}\label{SERE5015}
\Delta_1 = \left(1-\frac{{l_a}^2}{{i}^2 3!}\frac{\partial^2}{\partial x^2}+\frac{{l_a}^4}{{i}^4 5!}\frac{\partial^4}{\partial x^4}- ....\right)
\end{equation}
Following the similar approach as above, one can show that
\begin{equation}\label{SERE5016}
\Delta_4 = \left(1-\frac{{l_a}^2}{{i}^2 3!}\frac{\partial^2}{\partial x^2}+\frac{{l_a}^4}{{i}^4 5!}\frac{\partial^4}{\partial x^4}- ....\right)^4
\end{equation}
From the above details, it is reasonable to conclude that
\begin{equation}\label{SERE5017}
\Delta_m = \left(\Delta_1\right)^m
\end{equation}
These $\Delta_m$s can be used to derive some other useful unified nonlocal rational continuum equations very efficiently.

\subsubsection{Variational formalism: energy functional for general formulation}
The well-posedness of the problems, which involves the requirement of appropriate boundary conditions, can be obtained adopting variational formalism. Therefore, the governing equations with appropriate boundary conditions are to be derived from an appropriate elastic potential functional using Hamilton's principle. The Hamilton's principle for deformable conservative systems can be stated as
\begin{align}\label{SERE66011HP}
 \delta \int_{t_1}^{t_2} \left[W_k - (W_i + W_e)\right] \, dt = 0
\end{align}
where $W_k$ is the kinetic energy, $W_i$ is the elastic potential energy, and $W_e$ is the work done by external loads on the system. Here $t_1$, $t_2$ are the initial and final times, respectively. Observing the characteristic of the Eqs. (\ref{SERE508ex}) and (\ref{SERE53011ex}), the elastic potential energy functional for a nonlocal linear elastic system with small-strain condition can be assumed as
\begin{align}\label{SERE66011}
 W_i = \frac{1}{2} \int_V \sum_{p=1}^{N} D^{np}_{jklm} \epsilon^p_{jk} \epsilon^p_{lm} \, dv = \frac{1}{2} \int_V \sum_{p=1}^{N} \chi_p(\Delta_1) D^{p}_{jklm} \epsilon^p_{jk} \epsilon^p_{lm} \, dv; \;\; j,k,l,m=1,2,3
\end{align}
where $D^{np}_{jk}$ and $\epsilon^p_{jk}$ are the generalized nonlocal stiffness parameters which are functions of discrete dispersion differential operators and the local strains of the order $p$, respectively. The function $\chi_p(\Delta_1)$ is a function of $\Delta_1$, which takes account of the microstructural interaction of the continuum.  Here, the expression of $\chi_p(\Delta_1)$ depends upon the $p^{th}$ order of gradients of the independent displacement field variables in the strain field $\epsilon^p_{jk}$ and the degree of that $p^{th}$ order partial derivative term in the quantity $\epsilon^p_{jk} \epsilon^p_{lm}$. The nonlocal rational stress tensor $\sigma^{np}_{jk}$, associated with the $p^{th}$ order strains, can be defined as
\begin{align}\label{SERE66014}
 \sigma^{np}_{jk} = \frac{\partial}{\partial {\epsilon^p_{jk}}} \sum_{p=1}^{N} D^{np}_{jklm} \epsilon^p_{jk} \epsilon^p_{lm} = D^{np}_{jklm} {\epsilon^p_{lm}}
\end{align}
The local strains $\epsilon^p_{jk}$ are related to the local Cauchy stress tensor $\sigma^p_{jk}$ by the relation
\begin{align}\label{SERE66015}
 \sigma^p_{jk} = D^{p}_{jklm} {\epsilon^p_{lm}}
\end{align}
Using the formalism given in Eqs. (\ref{SERE66015})-(\ref{SERE66014}), one can obtain the expression of nonlocal rational stress   corresponding to the nonlocal rational rod equation (\ref{SERE508ex}) as
\begin{align}\label{SERE66016exact}
 \sigma^{n1}_{xx} = {\Delta_1}^2 E \epsilon^1_{xx} = {\Delta_1}^2 E \frac{\partial u}{\partial x} = \left(1-\frac{{l_a}^2}{12{i}^2}\frac{\partial^2}{\partial x^2}+\frac{{l_a}^4}{12{i}^4}\frac{\partial^4}{\partial x^4}- ....\right) \sigma^1_{xx}
\end{align}
Considering the criteria $|k_l l_a| < 1$, the exact expression in the Eq. (\ref{SERE66016exact}) can be approximated as
\begin{align}\label{SERE66017aprox}
 \sigma^{n1}_{xx} \simeq \left(1-{e_0}^2 {l_a}^2\frac{\partial^2}{\partial x^2}\right)^{-1} \sigma^1_{xx}
\end{align}
i.e., we have the equivalent differential form of the Eringen's nonlocal constitutive equation \citep{eringen1983differential,fernandez2016bending,raghu2016nonlocal} for 1D rod
\begin{align}\label{SERE66018aprox}
 \left(1-{e_0}^2 {l_a}^2\frac{\partial^2}{\partial x^2}\right)\sigma^{n} = \sigma
\end{align}
where $e_0$ is another nonlocal parameter other than the length scale parameter $l_a$. From which, it can be shown that the nonlocal constitutive equation for higher dimensions will be as
\begin{align}\label{SERE66019aprox}
 \left(1-{e_0}^2 {l_a}^2\nabla^2 \right)\sigma^{n} = \sigma
\end{align}
The above derivations clearly show that the popular strain gradient elasticity models \cite{toupin1962elastic,mindlin1964micro} and integral nonlocal elasticity models \citep{eringen1983differential} are special approximations of the novel unified nonlocal rational continuum model.

The next two sections describe the development procedures of some higher-order nonlocal rational continuum equations with improved dispersive characteristics and/or nonlocal properties in general.

\subsection{A nonlocal rational higher-order rod approximation of 2D harmonic lattices}
Any $2$D atomistic system can actually be idealized as $1$D higher-order rod and/or higher-order beam continuum with a very high accuracy for static and low frequency dynamic problems. However, these homogenized classical continuum models in the form of partial differential equations fail to represent the discrete atomistic systems satisfactorily in case of high frequency dynamic conditions. Therefore, with proper enrichment of these local continuum equations, the discrepancies between the continuum predictions and experimental observations can be resolved satisfactorily.

\begin{figure}[!h]
\centering
\includegraphics[scale=0.4]{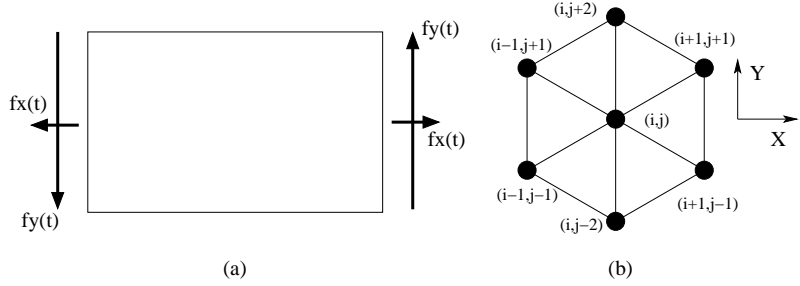}
\caption{Schematic of the $2$D triangular lattice system and unit cell: $(a)$ the rectangular 2D atomistic domain considered as a plane continuum; $(b)$ the unit cell with hexagonal nearest neighbor interactions.}
\label{2Dhcp}
\end{figure}

A simple $2$D triangular lattice system with hexagonal interaction (Fig. \ref{2Dhcp}) is considered for modeling  wave propagation in $2$D atomistic system. The triangular lattice system is chosen because it represents as either the basal plane of many hexagonal close-packed (hcp) lattice systems or the ($111$) plane of many face-centered cubic (fcc) lattice systems \citep{kittel2008sspbook}. There are an enormous number of other materials (not only Graphene, Graphyne, Germanene,  palladium, Molybdenum disulfide, etc.) that are 2D materials and are very interesting from the basic science and applications point of view. For many other 2D lattice systems, the formalism can be extended easily. Therefore, at first, the atomistic equations and dispersion relations for the $2$D system is described. Then, higher enriched order rod and beam approximations for 2D triangular lattice system are obtained simultaneously using atomistic information. The $n^{th}$ atom in $i^{th}$ column and $j^{th}$ row location in the $2$D lattice system, where $j=2n$ if $i$ is odd and $j=2n-1$ if $i$ is even, is indexed by $(i,n)\in \mathbb{Z}^2$. The position of the $(i,n)^{th}$ atom in the $X$-$Y$ plane at time $t$ is given by (Fig. \ref{2Dhcp})
\begin{align}\label{SEBE5021}
\left(
  \begin{array}{c}
    x \\
    y \\
  \end{array}
\right)_{i,n}= \left(
                \begin{array}{c}
                i r_x \\
                j r_y \\
                \end{array}
                \right) + \left(
                            \begin{array}{c}
                            u_{i,n}(t) \\
                            v_{i,n}(t) \\
                            \end{array}
                            \right) \in \mathbb{R}^2
\end{align}
Here $r_x=\frac{\sqrt{3}r_0}{2}$, $r_y=\frac{r_0}{2}$ (Fig. \ref{2Dhcp}), where $r_0$ is the equilibrium atomic separation and $(u_{i,n}(t), v_{i,n}(t))^T = d_{i,n}(t)$ are out-of-equilibrium atomic displacements along the respective directions in the $X$-$Y$ plane. Considering nearest-neighbor interactions (Fig. \ref{2Dhcp}) at first, the Hamiltonian for the system can be written as \citep{friesecke2003geometric}
\begin{align}\label{SEBE5022}
H = \displaystyle \sum_{i,n} (\frac{1}{2m_a}\texttt{|}p_{i,n}\texttt{|}^2 + V_1(\texttt{|}(e_1+e_2) + d_{i+1,j+1}- d_{i,n}\texttt{|}) \;\;\;\;\;\;\;\;\;\;\;\;\;\;\;\; \nonumber\\
 + V_1(\texttt{|}(e_1-e_2) + d_{i+1,j-1}- d_{i,n}\texttt{|})+ V_1(\texttt{|}e_2 + d_{i,j+2}- d_{i,n}\texttt{|}))
\end{align}
Here $\texttt{|}.\texttt{|}$ denotes the Euclidian norm on $\mathbb{R}^2$, $p_{i,n}$ denotes the momentum vector of the $(i,n)^{th}$ atom, $e_1$ and $e_2$ are the lattice basis vectors $(\frac{\sqrt{3}r}{2},0)$ and $(0,\frac{r}{2})$, respectively. $V_1$ denotes the interatomic potential function of any arbitrary type. However, in the present work, we consider potential incorporating stretching mode only. Although, the nearest-neighbor interaction is assumed initially, however, this formalism can be easily extended for non-nearest neighbor interactions. The potential $V_1$ may be any arbitrarily differentiable function, but for continuum approximations, it is reasonable to consider equivalent harmonic potential of the form
\begin{align}\label{SEBE5023}
V_1(\texttt{|}z\texttt{|}) = \frac{S_1}{2}(\texttt{|}z\texttt{|} - r)^2
\end{align}
where $S_1$ is the equivalent harmonic spring constant and $\texttt{|}z\texttt{|}$ denotes the distance of separation between any two nearest neighbor atoms. For any other nonlinear potentials, the constants of equivalent harmonic springs are obtained from the curvature of the nonlinear potential. The equations of motion of the $(i,n)^{th}$ atom can be written as
\begin{align}\label{SEBE5024}
m_a \ddot{d}_{i,n} = -\frac{\partial H}{\partial {d}_{i,n}}
\end{align}
Expanding Eq. (\ref{SEBE5024}), the equations of motion can be written as
\begin{align}\label{SEBE5025}
m_a \ddot{d}_{i,n} = -\{-f_1(2e_2 + d_{i,j+2} - d_{i,n}) + f_1(2e_2 + d_{i,n} - d_{i,j-2}) \;\;\;\;\;\;\;\;\;\;\; \;\;\;\;\;\;\;\; \nonumber\\
-f_2((e_1+e_2) + d_{i+1,j+1} - d_{i,n}) + f_2((e_1+e_2) + d_{i,n} - d_{i-1,j-1}) \nonumber\\
-f_2(r(e_1-e_2) + d_{i+1,j-1} - d_{i,n}) + f_2((e_1-e_2) + d_{i,n} - d_{i-1,j+1})
\}
\end{align}
Where the forcing terms are defined as $f_1(z)={V_1}^{\prime}(\texttt{|}z\texttt{|}) \frac{z}{\texttt{|}z\texttt{|}}$. Here, the Hamiltonian equations of motion are still nonlinear, due to the frame-indifference of the interatomic forces \citep{friesecke2003geometric}. This is analogous to the geometric nonlinearity in continuum elasticity. Therefore, Eq. (\ref{SEBE5025}) must be linearized for studying the phonon dispersion characteristic of the 2D lattice. The linearized equations can be obtained from Eq. (\ref{SEBE5025}) as
\begin{align}\label{SEBE5026}
m_a \ddot{d}_{i,n} = \{ M_1(d_{i+1,j+1} - 2d_{i,n} + d_{i-1,j-1}) \;\;\;\;\;\;\;\;\;\;\; \;\;\;\;\;\;\;\;\nonumber\\
+ M_2(d_{i+1,j-1} - 2d_{i,n} + d_{i-1,j+1}) + M_3(d_{i,j+2} - 2d_{i,n} + d_{i,j-2})
\}
\end{align}
Where $M_1$, $M_2$, and $M_3$ are $2\times 2$ matrices as given below
\begin{equation}\label{SEBE5027}
M_1 = S_1\left[
          \begin{array}{cc}
            \frac{3}{4} & \frac{\sqrt{3}}{4}\\
            \frac{\sqrt{3}}{4} & \frac{1}{4}\\
          \end{array}
         \right], \; \;
M_2 = S_1\left[
          \begin{array}{cc}
            \frac{3}{4} & -\frac{\sqrt{3}}{4}\\
            -\frac{\sqrt{3}}{4} & \frac{1}{4}\\
          \end{array}
         \right], \; \;
M_3 = S_1\left[
          \begin{array}{cc}
            0 & 0\\
            0 & 1\\
          \end{array}
         \right] \\
\end{equation}

Consider the $2$D atomistic system as a deep axial waveguide under loading in X-direction only (Fig. \ref{2Dhcp}). Let the wave propagate in the X-direction as considered in \cite{friesecke2003geometric}. Motivated by the atomistic solutions of Eq. (\ref{SEBE5025}), the approximate continuum deformation field variables are assumed to be of the form
\begin{align}\label{SEBE5028}
u(x,y,t)=u_0(x,t), \; \; \;\; \; v(x,y,t)=y\psi_0 (x,t)
\end{align}
where $u_0(x,t)$ and $\psi_0 (x,t)$ are displacement field variables corresponding the neutral axis along X-direction and $\psi_0 (x,t)= \frac{\partial v}{\partial y}|_{y=0}$ takes account of the Poisson's effect. The local strain fields corresponding to the field variables in Eq. (\ref{SEBE5028}) are computed as
\begin{align}\label{SEBE50282}
\epsilon_{xx}=\frac{\partial u_0(x,t)}{\partial x}, \; \; \epsilon_{yy}= \psi_0 (x,t), \;\; \; \gamma_{xy}=y\frac{\partial \psi_0 (x,t)}{\partial x}
\end{align}
Using Hamilton's principle, two coupled equations of motion are obtained for the assumed field variables in Eq. (\ref{SEBE5028}) as \citep{patra2014spectral,gopalakrishnan2007spectral}
\begin{align}\label{SEBE5029}
C_{11}A \frac{\partial^2 u_0}{\partial x^2} + C_{12}A \frac{\partial \psi_0}{\partial x} = \rho A \frac{\partial^2 u_0}{\partial t^2}, \nonumber\\
C_{66}I \gamma_1 \frac{\partial^2 \psi_0}{\partial x^2} - C_{11}A \psi_0 - C_{12}A \frac{\partial u_0}{\partial x} = \rho I \gamma_2 \frac{\partial^2 \psi_0}{\partial t^2}
\end{align}
Here, $I$, $\rho$, and $A$ are the second moment of area of cross-section, the mass density, and the area of cross-section of the 2D atomistic lamina, respectively. In Eq. (\ref{SEBE5029}), $\gamma_1$ and $\gamma_2$ are adjustable parameters \citep{doyle1997wave, gopalakrishnan2000deep} and $C_{11}$, $C_{12}$, and $C_{66}$ are stiffness parameters, respectively. For generality, the continuum parameters $C_{11}$, $C_{12}$ can be obtained using the following relation \citep{kittel2008sspbook}
\begin{align}\label{SEBE5030}
 C_{11} = \frac{1}{A_a h}\frac{\partial^2 U_a}{\partial {\epsilon^2}_{xx}}|_{\epsilon_{xx}=0} \;\;\;\;\; C_{12} = \frac{1}{A_a h}\frac{\partial^2 U_a}{\partial \epsilon_{yy} \partial \epsilon_{xx}}|_{\epsilon_{xx}=0,\epsilon_{yy}=0}
\end{align}
where $U_a$ is potential energy per atom in a unit cell under uniform tensile deformation, $A_a=\frac {\sqrt{3}}{2} r^2$ is the effective surface area occupied by the atom in the X-Y plane, $h$ is the thickness of the $2$D atomistic lamina and $\epsilon_{xx}=(u_{i+1,j+1}-u_{i,j})/r_x$, $\epsilon_{yy}=(v_{i+1,j+1}-v_{i,j})/r_y$ are strains, respectively. From Eq. (\ref{SEBE5030}), the parameters for the triangular lattice are obtained as $C_{11}=\frac {3\sqrt{3}}{4} \frac {S_1}{h}$, $C_{12}=\frac {\sqrt{3}}{4} \frac {S_1}{h}$. The parameter $C_{66}= \frac {\sqrt{3}}{4} \frac {S_1}{h}$ is obtained from $C_{11}$ and $C_{12}$. The associated boundary conditions for Eq. (\ref{SEBE5029}) are specified as
\begin{align}\label{SEBE5031}
F = C_{11}A \frac{\partial u_0}{\partial x} + C_{12}A \psi_0, \; \; \; Q = C_{66}I \gamma_1  \frac{\partial \psi_0}{\partial x}
\end{align}
The equations of motion in (\ref{SEBE5029}) are the so called local Mindlin-Herrmann rod (LMHR) equations \citep{mindlin1951one, doyle1997wave}. This classical Mindlin-Herrmann rod (LMHR) model (Eq. (\ref{SEBE5029})) shows poor dispersive characteristics as compared to its atomistic counterpart (Eq. (\ref{SEBE5025})), because of the homogenization procedure involved in the formulation of continuum equations. To overcome this snag, the homogenized $\textbf{L}$ operator in Eq. (\ref{SEBE5029}) is to be replaced with the nonlocal $\bar{\textbf{L}}$ operator containing appropriate $\Delta_m$s. However, the nonlocal elastic potential functional form corresponding to the unified nonlocal rational Mindlin-Herrmann rod (NRMHR) equations can be written according to Eq. (\ref{SERE66011}) as
\begin{align}\label{SEBE5031potential}
\small
W_i = \frac{1}{2} \int_V [ C_{11} {\Delta_1}^2 \left(\frac{\partial u_0(x,t)}{\partial x}\right)^2 + C_{12} {\Delta_0} {\Delta_1} \psi_0 (x,t) \frac{\partial u_0(x,t)}{\partial x} \nonumber\\ + C_{22} {\Delta_0}^2 (\psi_0 (x,t))^2 + C_{66} y^2 {\Delta_1}^2 \left(\frac{\partial \psi_0 (x,t)}{\partial x}\right)^2 ] dv
\end{align}
Therefore, the novel nonlocal rational Mindlin-Herrmann rod (NRMHR) equations are derived using Hamilton's variational principle considering the elastic potential functional form given in Eq. (\ref{SEBE5031potential}) as
\begin{align}\label{SEBE5032}
C_{11}A \Delta_2 \frac{\partial^2 u_0}{\partial x^2} + C_{12}A \Delta_1 \frac{\partial \psi_0}{\partial x} = \rho A \frac{\partial^2 u_0}{\partial t^2}, \nonumber\\
C_{66}I \gamma_1 \Delta_2 \frac{\partial^2 \psi_0}{\partial x^2} - C_{11}A \psi_0 - C_{12}A \Delta_1 \frac{\partial u_0}{\partial x} = \rho I \gamma_2 \frac{\partial^2 \psi_0}{\partial t^2}
\end{align}
and associated boundary conditions for Eq. (\ref{SEBE5032}) are obtained variationally as
\begin{align}\label{SEBE5033}
F = C_{11}A (\Delta_1)^2 \frac{\partial u_0}{\partial x} + \Delta_1 C_{12}A \psi_0, \; \; \; Q = C_{66}I \gamma_1 (\Delta_1)^2 \frac{\partial \psi_0}{\partial x}
\end{align}
where $\Delta_m = (\Delta_1)^m$ and $\Delta_1 = (1-\frac{{r_x}^2}{{i}^2 3!}\frac{\partial^2}{\partial x^2}+\frac{{r_x}^4}{{i}^4 5!}\frac{\partial^4}{\partial x^4}- ....)$ for the present lattice configuration.

For studying the dispersion relation of the new nonlocal rational Mindlin-Herrmann rod equations, we proceed as follows. Since there are two independent variables $u_0$ and $\psi_0$, the homogeneous traveling wave solutions are assumed of the form
\begin{align}\label{SEBE5034}
u_0 = \hat{u}_0 e^{-i_m(k x - \omega t)}, \;\;\;\; \psi_0 = \hat{\psi}_0 e^{-i_m(k x - \omega t)}
\end{align}
where $i_m = \sqrt{-1}$ and $k$ is the wavenumber and $\omega$ is the circular frequency. Substituting these solutions into Eq. (\ref{SEBE5032}) gives
\begin{equation}\label{SEBE5035}
  \left[
  \begin{array}{cc}
  -C_{11}A (\frac{\sin \frac{kr_x}{2}}{\frac{r_x}{2}})^2 + \omega^2 \rho A & -i_mC_{12}A(\frac{\sin \frac{kr_x}{2}}{\frac{r_x}{2}}) \\
  i_mC_{12}A(\frac{\sin \frac{kr_x}{2}}{\frac{r_x}{2}}) & -C_{66}I\gamma_1(\frac{\sin \frac{kr_x}{2}}{\frac{r_x}{2}})^2 - C_{11}A + \omega^2 \rho I\gamma_2\\
  \end{array}
  \right]\left(
         \begin{array}{c}
         \hat{u}_0 \\
         \hat{\psi}_0\\
         \end{array}\right)
         =\left(
         \begin{array}{c}
         0 \\
         0 \\
         \end{array}\right)
\end{equation}
Setting the determinant of the above matrix to zero gives the characteristic equation in the general case and its solution gives four possible modes as
\begin{align}\label{SEBE5036}
k(\omega)_{1,3} = \frac{2}{r_x}\sin^{-1} (\pm \frac{r_x}{2}\left[ -\frac{b}{2a} + \sqrt{\frac{b^2}{4a^2} - \frac{c}{a}}\right]^{\frac{1}{2}})\nonumber\\
k(\omega)_{2,4} = \frac{2}{r_x}\sin^{-1} (\pm \frac{r_x}{2}\left[ -\frac{b}{2a} - \sqrt{\frac{b^2}{4a^2} - \frac{c}{a}}\right]^{\frac{1}{2}})
\end{align}
Where $a = C_{11}C_{66} AI\gamma_1$, $b = (A^2 C_{11}^2 - A^2 C_{12}^2 - \omega^2 C_{11} \rho AI\gamma_2 - \omega^2 C_{66} \rho AI\gamma_1)$, and $c = (-\omega^2C_{11} \rho A^2 + \omega^4 \rho^2 AI\gamma_2)$. These are the two spectrum relations of the NRMHR equations. The circular frequency $\omega$ is actually a proper function of the wavenumber $k$. Here, however, it is considered that the frequency $\omega$ will have real values only and $k$ might be real as well as complex. Therefore, the wave number $k$ is expressed as a function of the circular frequency of vibration $\omega$. This consideration is particularly advantageous for the computations in frequency space. The propagating group speed-frequency dependance can be obtained from these equations as
\begin{align}\label{SEBE5037}
C_{gc1} = \frac{\partial \omega}{\partial k_1}\nonumber\\
C_{gc2} = \frac{\partial \omega}{\partial k_2}
\end{align}
Here, $C_{gc1}$ and $C_{gc2}$ are propagating longitudinal and contractional wave speeds along the X-direction of the 2D lamina, respectively. These are the continuum predictions of two propagating group velocities by the NRMHR equations.

To study the dispersion characteristics of the actual 2D atomistic lamina for the similar type of traveling wave solution consideration, one can assume the atomistic displacement fields of the form
\begin{align}\label{SEBE5038}
u_{i,n} = \hat{u}_0 e^{-i_m(k ir_x - \omega t)}, \;\;\;\; v_{i,n} = y_n\hat{\psi}_0 e^{-i_m(k ir_x - \omega t)} = (j-j_0)r_y\hat{\psi}_0 e^{-i_m(k ir_x - \omega t)}
\end{align}
where $j_0$ is row number of the $y=0$ axis along X-direction of the lamina. Substituting these solutions in the linearized atomistic equations (\ref{SEBE5026}) for the 2D lamina and after some manipulation, the eigenvalue equation is obtained as
\begin{equation}\label{SEBE5039}
  \left[
  \begin{array}{cc}
  \frac{6S_1}{m_a} \sin^2 \frac{kr_x}{2} - \omega^2 & i_m\sqrt{3}r_y \sin kr_x \\
  0 & \frac{2S_1}{m_a} \sin^2 \frac{kr_x}{2} - \omega^2\\
  \end{array}
  \right]\left(
         \begin{array}{c}
         \hat{u}_0 \\
         \hat{\psi}_0\\
         \end{array}\right)
         =\left(
         \begin{array}{c}
         0 \\
         0 \\
         \end{array}\right)
\end{equation}
Setting the determinant of the above matrix to zero gives the characteristic equation in the general case and its solution gives atomistic dispersion relation as
\begin{align}\label{SEBE50310}
k(\omega)_{1,3} = \frac{2}{r_x}\sin^{-1} (\pm \omega \sqrt{\frac{m_a}{2S_1}})\nonumber\\
k(\omega)_{2,4} = \frac{2}{r_x}\sin^{-1} (\pm \omega \sqrt{\frac{m_a}{6S_1}})
\end{align}
Here also, for the atomistic dispersion relation, we have expressed $k$ as a function of $\omega$ to keep parity with the previous continuum expression of the same. The two propagating group speeds are obtained from the above two expressions as
\begin{align}\label{SEBE50311}
C_{ga1} = \frac{\partial \omega}{\partial k_1} = \frac{r_x}{2} (\sqrt{\frac{2S_1}{m_a}}) \cos \frac{k_1r_x}{2} = \frac{r_x}{2} (\sqrt{\frac{2S_1}{m_a} - \omega^2}) \nonumber\\
C_{ga2} = \frac{\partial \omega}{\partial k_2} = \frac{r_x}{2} (\sqrt{\frac{6S_1}{m_a}}) \cos \frac{k_2r_x}{2} = \frac{r_x}{2} (\sqrt{\frac{6S_1}{m_a} - \omega^2})
\end{align}
Here, $C_{ga1}$ and $C_{ga2}$ are the atomistic predictions of propagating group speeds corresponding to the stretching and contraction modes, respectively.

In the above derivation of nonlocal rational Mindlin-Herrmann rod (NRMHR) equations as an equivalent continuum model of the actual 2D atomistic lattice, only nearest neighbor interactions are considered initially. For the consideration of any non-nearest neighbor interactions, the model can be extended variationally taking the contribution of higher order strains in the energy functional (Eq. (\ref{SERE66011})).

\subsection{A nonlocal rational higher-order beam approximation of 2D harmonic lattices}
Now consider the $2$D atomistic system as a flexural waveguide under loading in Y-direction only, where the wave is propagating in the X-direction. Then, the atomistic solutions motivate the approximate continuum field variables to be of the form
\begin{align}\label{SEBE5041b}
u(x,y,t)= - y\phi_0 (x,t), \; \; \;\; \; v(x,y,t) = v_0 (x,t)
\end{align}
where $\phi_0 (x,t)$ is the angle due to pure bending. This is same as the assumption made in the classical Timoshenko beam theory \citep{timoshenko1921lxvi}. Therefore, using Hamilton's principle, one can obtain two coupled equations of the classical Timoshenko beam (LTB) theory as \citep{gopalakrishnan1992matrix}
\begin{align}\label{SEBE5042b}
G A K_1 \left[\frac{\partial^2 v_0}{\partial x^2} - \frac{\partial \phi_0}{\partial x}\right] = \rho A \frac{\partial^2 v_0}{\partial t^2}, \nonumber\\
EI \frac{\partial^2 \phi_0}{\partial x^2} + GAK_1 \left[\frac{\partial v_0}{\partial x} - \phi_0 \right] = \rho I K_2 \frac{\partial^2 \phi_0}{\partial t^2}
\end{align}
Here $G$ and $E$ are shear modulus and Young's modulus, respectively, where the continuum parameters $G = C_{66}$ and $E=\frac {\sqrt{3}}{2} \frac {S_1}{h}$ is obtained using $C_{11}$ and $C_{12}$ with plain stress consideration. The other important parameter is Poisson's ratio $\sigma=\frac{1}{3}$ for the same consideration of plain stress. $K_1$ and $K_2$ are adjustable parameters \citep{doyle1990spectrally,gopalakrishnan1992matrix,dong2010much} that are required to approximate shear stress behavior across the beam depth and rotational inertias, respectively. The associated boundary conditions to be satisfied for Eq. (\ref{SEBE5042b}) are
\begin{align}\label{SEBE5043b}
V = GAK_1 \left[\frac{\partial v_0}{\partial x} - \phi_0 \right] = - EI \frac{\partial^2 \phi_0}{\partial x^2} +  \rho I K_2 \frac{\partial^2 \phi_0}{\partial t^2}, \;\;\;\; \; M = EI \frac{\partial \phi_0}{\partial x}
\end{align}
Following similar technique adopted as in the case of NRMHR formulation, we obtain the nonlocal rational Timoshenko beam equation (NRTB) as
\begin{align}\label{SEBE5044b}
G A K_1 \left[\Delta_2\frac{\partial^2 v_0}{\partial x^2} - \Delta_1\frac{\partial \phi_0}{\partial x}\right] = \rho A \frac{\partial^2 v_0}{\partial t^2}, \nonumber\\
EI\Delta_2 \frac{\partial^2 \phi_0}{\partial x^2} + GAK_1 \left[\Delta_1 \frac{\partial v_0}{\partial x} - \phi_0\right] = \rho I K_2 \frac{\partial^2 \phi_0}{\partial t^2}
\end{align}
Similarly, the associated enriched boundary conditions to be satisfied for Eq. (\ref{SEBE5044b}) are
\begin{align}\label{SEBE5045b}
V = GAK_1 \Delta_1 \left[\Delta_1 \frac{\partial v_0}{\partial x} - \phi_0\right] = \Delta_1 (- EI \Delta_2 \frac{\partial^2 \phi_0}{\partial x^2} +  \rho I K_2 \frac{\partial^2 \phi_0}{\partial t^2}), \;\;\;\; \; M = EI \Delta_2 \frac{\partial \phi_0}{\partial x}
\end{align}
Since there are two independent variables $v_0$ and $\phi_0$, the homogeneous solutions are assumed as
\begin{align}\label{SEBE5046b}
v_0 = \hat{v}_0 e^{-i_m(k_b x - \omega t)}, \;\;\;\; \phi_0 = \hat{\phi}_0 e^{-i_m(k_b x - \omega t)}
\end{align}
Here $k_b$ is the wave number of beam vibration. Substituting these solutions into Eq. (\ref{SEBE5044b}) gives
\begin{equation}\label{SEBE5047b}
  \left[
  \begin{array}{cc}
  GAK_1(\frac{\sin \frac{kr_x}{2}}{\frac{r_x}{2}})^2 - \omega^2 \rho A & -i_mGAK_1(\frac{\sin \frac{kr_x}{2}}{\frac{r_x}{2}}) \\
  i_mGAK_1(\frac{\sin \frac{kr_x}{2}}{\frac{r_x}{2}}) & EI(\frac{\sin \frac{kr_x}{2}}{\frac{r_x}{2}})^2 + GAK_1 - \omega^2 \rho IK_2\\
  \end{array}
  \right]\left(
         \begin{array}{c}
         \hat{v}_0 \\
         \hat{\phi}_0\\
         \end{array}\right)
         =\left(
         \begin{array}{c}
         0 \\
         0 \\
         \end{array}\right)
\end{equation}
Setting the determinant of the above matrix to zero gives the characteristic equation in the general case as
\begin{align}\label{SEBE5048b}
(GAK_1EI)\vartheta^4 - (\omega^2GAK_1 \rho I K_2 + \omega^2EI \rho A)\vartheta^2 - (GAK_1 - \omega^2 \rho IK_2)\omega^2 \rho A = 0
\end{align}
where $\vartheta = (\sin \frac{kr_x}{2}/\frac{r_x}{2})$. Solution of Eq. (\ref{SEBE5048b}) gives four possible modes as
\begin{align}\label{SEBE5049b}
k_b(\omega)_{1,3} =\frac{2}{r_x}\sin^{-1} \left(\pm \frac{r_x}{2}\left[ \frac{1}{2}\left(\frac{1}{c_s} + \frac{Q}{c_q}\right)\omega^2 + \omega \sqrt{\frac{1}{c_q} + \frac{1}{4}\left(\frac{1}{c_s} - \frac{Q}{c_q}\right)^2 \omega^2}\right]^{\frac{1}{2}}\right)\nonumber\\
k_b(\omega)_{2,4} =\frac{2}{r_x}\sin^{-1} \left(\pm \frac{r_x}{2}\left[ \frac{1}{2}\left(\frac{1}{c_s} + \frac{Q}{c_q}\right)\omega^2 - \omega \sqrt{\frac{1}{c_q} + \frac{1}{4}\left(\frac{1}{c_s} - \frac{Q}{c_q}\right)^2 \omega^2}\right]^{\frac{1}{2}}\right)
\end{align}
where, the constants $c_s=GAK_1/\rho A$, $c_q=EI/\rho A$, and $Q=\rho IK_2/\rho A$. The propagating group speed-frequency dependance for this enriched Timoshenko beam can be obtained from these equations (\ref{SEBE5049b}) as
\begin{align}\label{SEBE50410b}
C_{bc1} = \frac{\partial \omega}{\partial k_{b1}}\nonumber\\
C_{bc2} = \frac{\partial \omega}{\partial k_{b2}}
\end{align}
For any non-nearest neighbor interactions, the nonlocal rational Timoshenko beam (NRTB) model can be developed variationally using proper  energy functional $W_i$, which incorporates the contribution of higher-order strains. These nonlocal rational continuum equations are easy to solve in the frequency domain and can be solved semi-analytically in the NLSFEM framework. Accuracy of these enriched equations is presented in the next section through some motivating analysis.

\section{Analytical and numerical results}
In this section, some analytical and numerical results are presented to show the motivating capabilities of these novel nonlocal rational continuum equations. Analytical solutions of the nonlocal rational continuum models are compared with the actual atomistic predictions. Most interestingly, these elegant nonlocal continuum equations are easy solvable in the NLSFEM framework. Here, the numerical solutions of these continuum equations using NLSFEM are compared with the same solutions obtained by molecular dynamics (MD). The NLSFEM, and MD techniques are described in detail in Ref. \citep{patra2014spectral}.

\begin{figure}[!h]
    \label{fig:subfigures}
    \begin{center}
        \subfigure[]{%
            \label{KnvsWn}
            \includegraphics[width=0.4\textwidth]{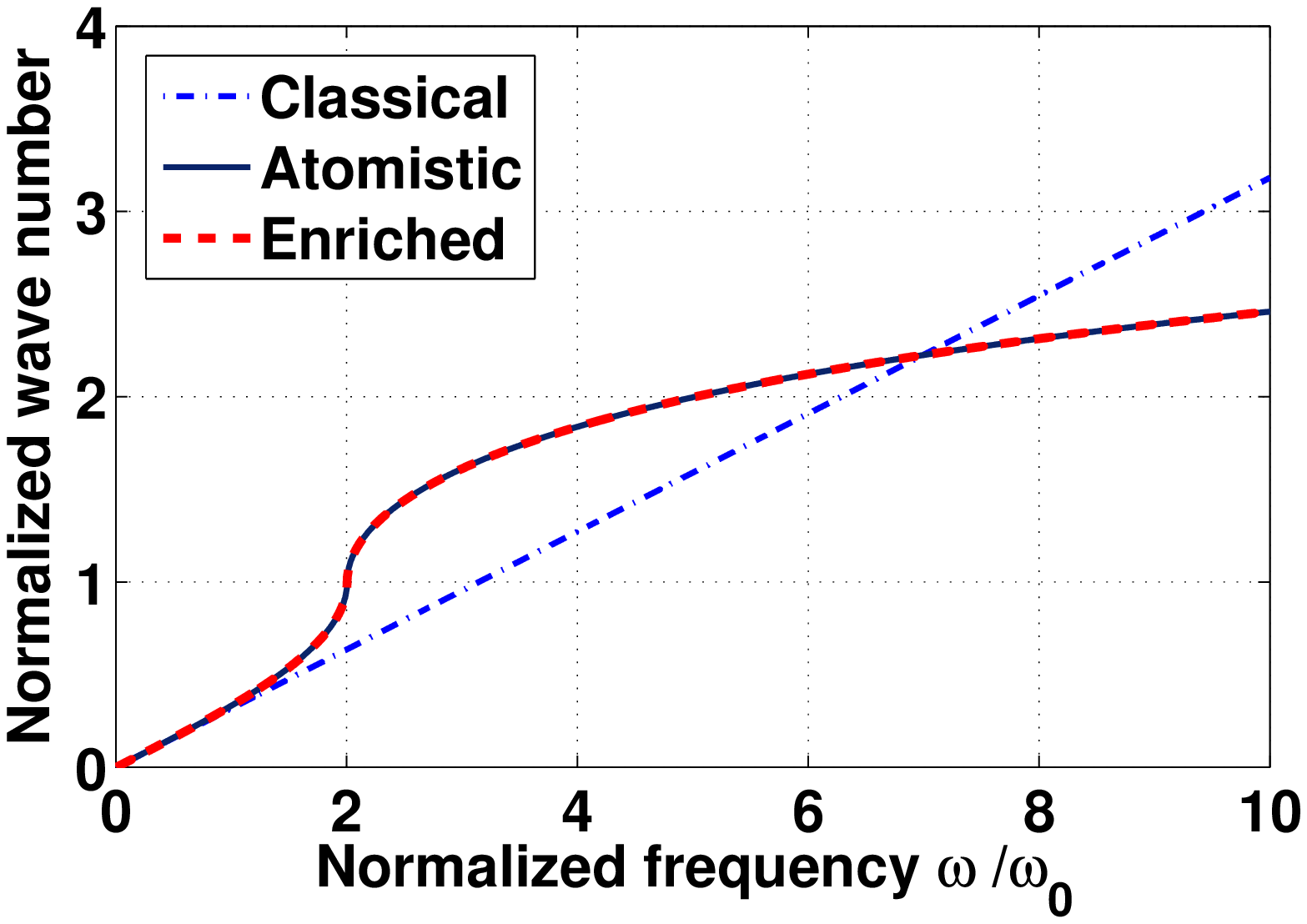}
        }%
        \subfigure[]{%
           \label{CgvsWn}
           \includegraphics[width=0.4\textwidth]{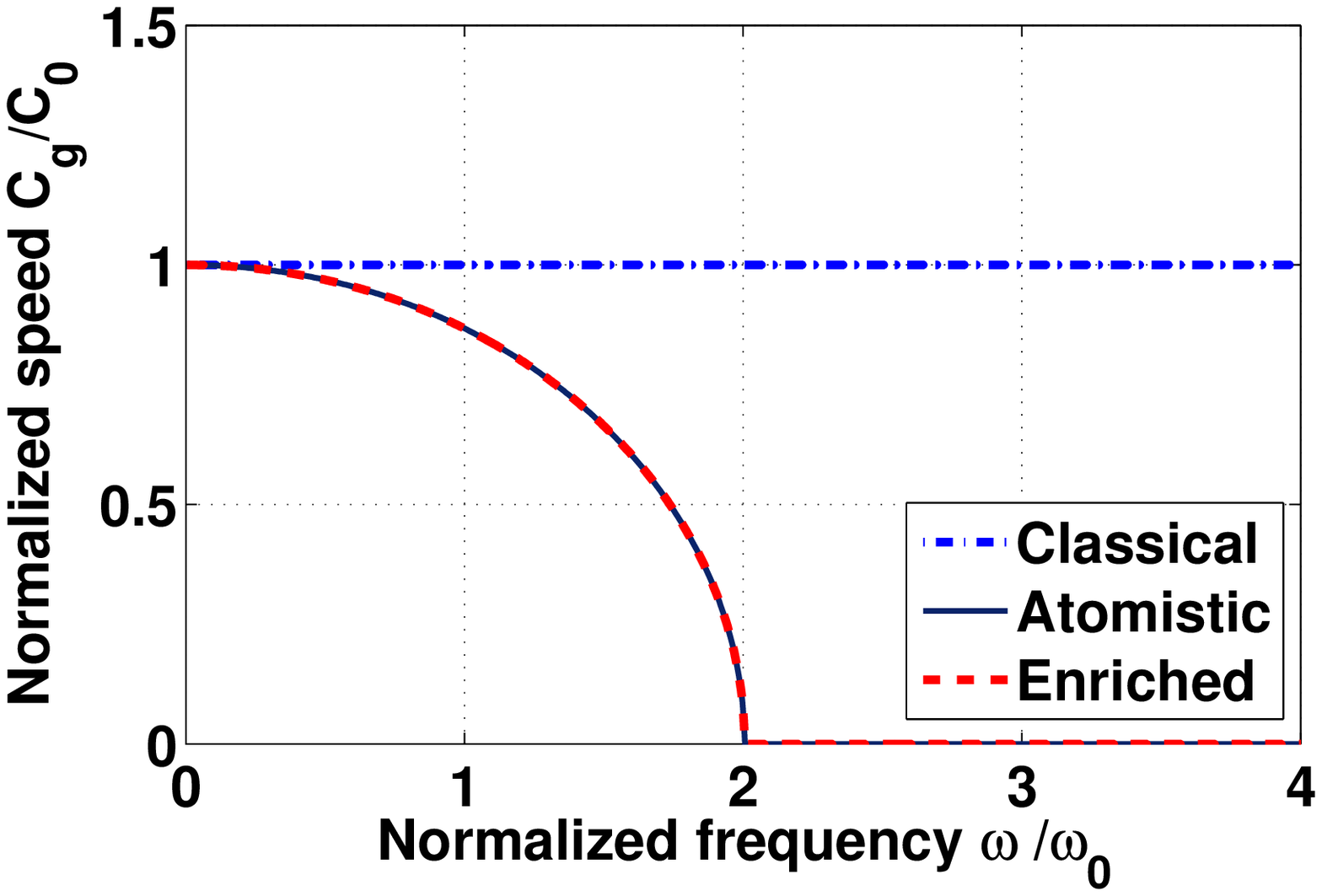}
        }
        \\ 
%
    \end{center}
    \caption{%
        Comparison of dispersion plots predicted by the atomistic equation (\ref{SERE503}), the enriched nonlocal rational rod equation (\ref{SERE508ex}), and the classical rod equation (\ref{SERE5302}):
        $(a)$ $(Re(k_l)-Im(k_l))/k_0$ vs $\omega/\omega_0$ plot; $(b)$ $C_g/C_0$ vs $\omega/\omega_0$ plot.
        Here $k_0 = \pi/l_a$, $\omega_0 = \sqrt{\frac{s_a}{m_a}}$, and $C_0=l_a\omega_0$.
        }%
     \label{Har1DNN}
\end{figure}
\subsection{Dynamics of $1$D harmonic lattice with nearest and non-nearest neighbor interactions}
For studying the capabilities of newly derived nonlocal 1D rod equations ((\ref{SERE508ex}), (\ref{SERE53011ex})) a harmonic lattice in one space dimension shown in Fig. \ref{harmoniclatt} is considered. The harmonic lattice consists of $2001$ atoms of mass $m_a=1\times 10^{-3}$ kg with an equilibrium interatomic spacing of $l_a=2.5\times 10^{-4}$ m, first of which from the left is considered to be fixed. Here, two ranges of interaction forces between the neighboring atoms are considered. Case-1 considers harmonic interaction forces only between the nearest neighbors with a spring stiffness $s_a=5\times 10^{11}$ N/m. This discrete lattice is exactly represented by the newly developed nonlocal rational rod (NRR) equation (\ref{SERE508ex}). In case-2, next-nearest neighbor interaction is considered with a spring stiffness $s_{a2}$ of variable magnitudes to study its effect on the changes in dispersive characteristics of the 1D lattice. In this case, the correct equivalent nonlocal rod (NRR) equation is Eq. (\ref{SERE53011ex}). First, we present the analytical phonon dispersion relations for both the cases. Then, the dynamic responses for case-1 are computed numerically using NLSFEM and MD for the comparative study. In the present analysis, such values of the lattice parameters (i.e., $m_a$,$l_a$,$s_a$, and $s_{a2}$) are chosen to reduce the load of MD computation. However, this doesn't deteriorate the qualitative behavior of the $1$D lattice.
\begin{figure}[!h]
    \label{fig:subfigures}
    \begin{center}
        \subfigure[]{%
            \label{KnvsWnNNN}
            \includegraphics[width=0.4\textwidth]{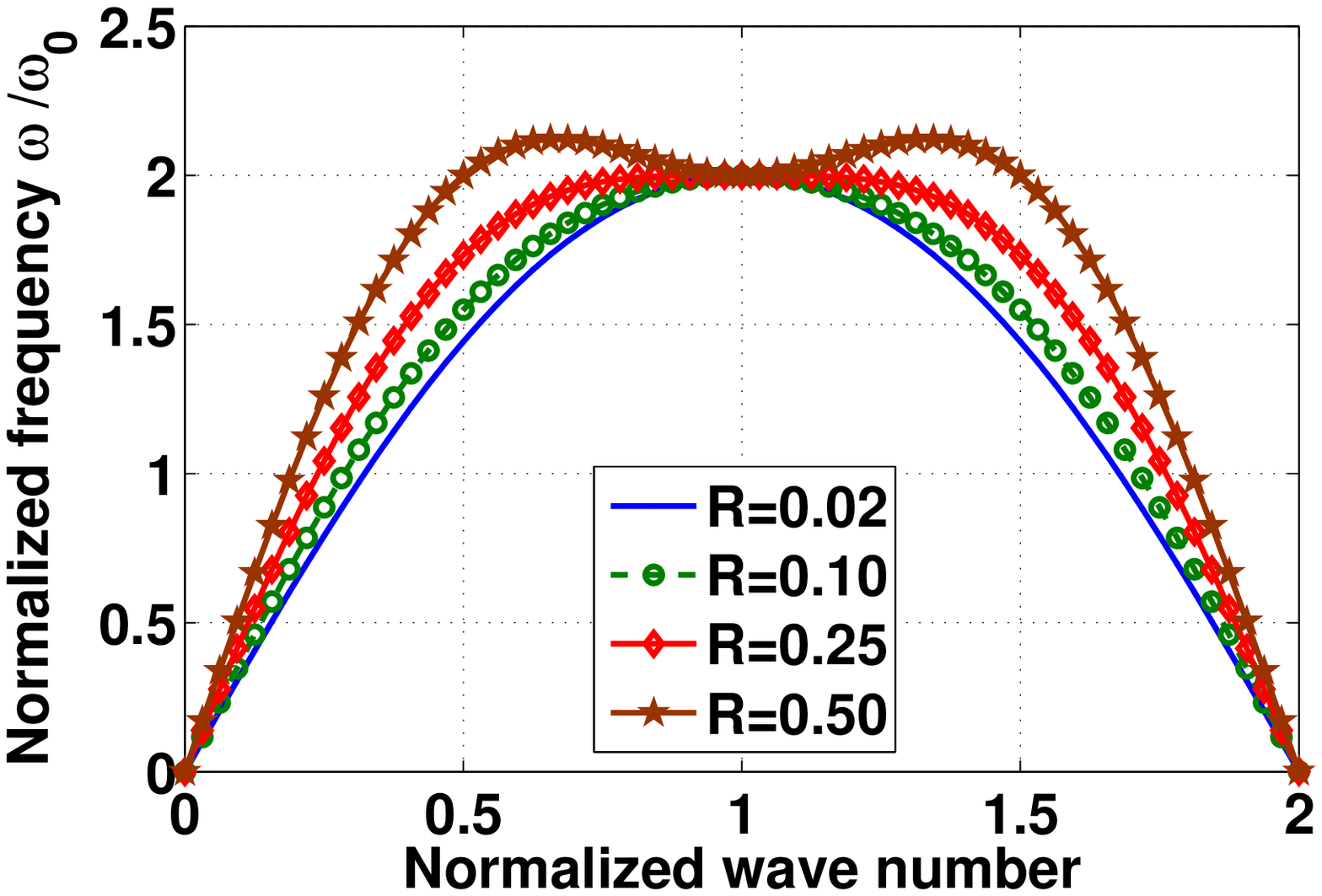}
        }%
        \subfigure[]{%
           \label{CgvsWnNNN}
           \includegraphics[width=0.4\textwidth]{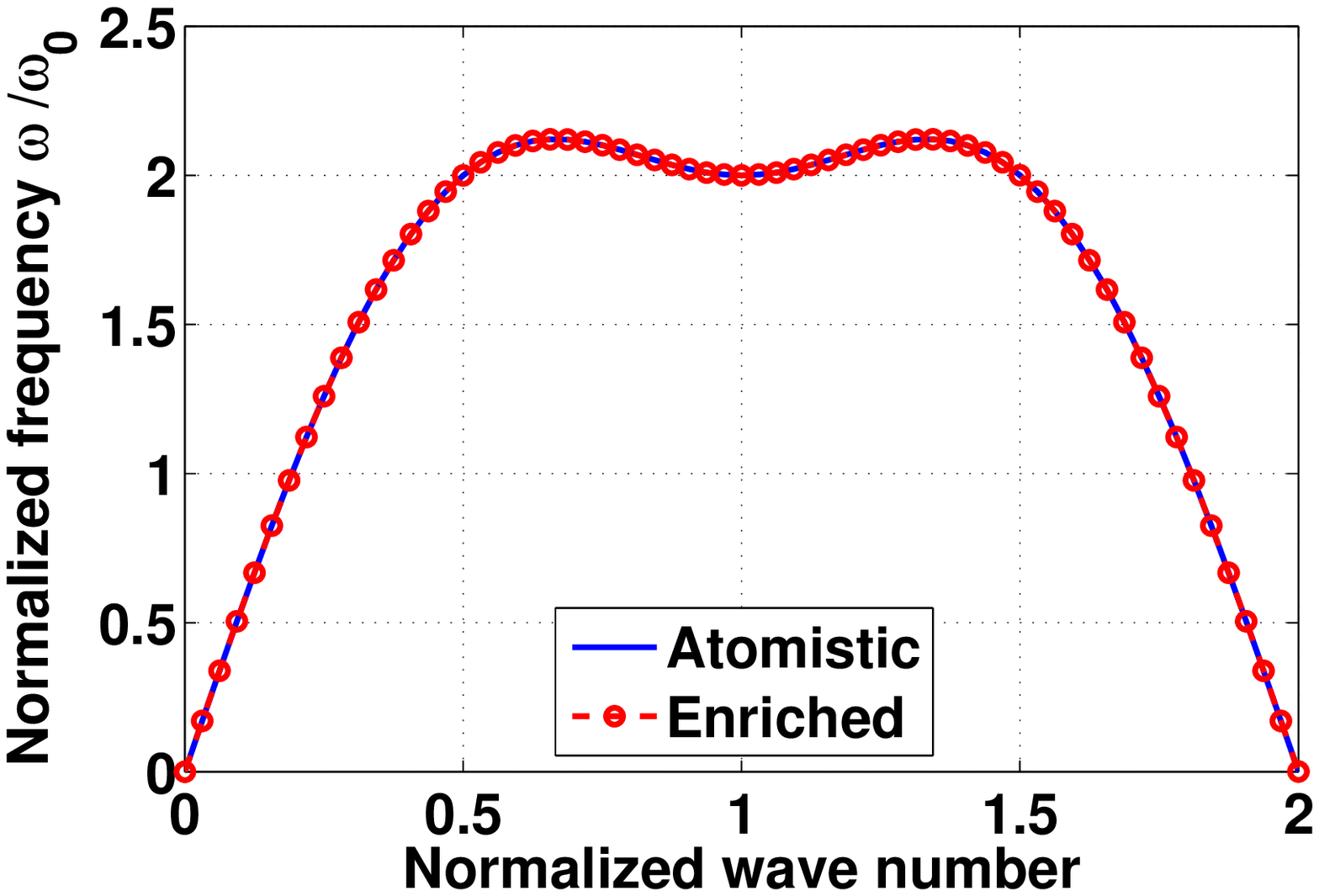}
        }
        \\ 
%
    \end{center}
    \caption{%
        Variation of dispersion plots with the variation of stiffness ratio $R$ for a $1$D harmonic lattice with next-nearest neighbor interactions, obtained using the enriched NRR equation (\ref{SERE53011ex}) and  its atomistic counterpart. The region of normalized wavenumber $\leq 1$ corresponds to first Brillouin zone, i.e., the limiting range of wavenumber $k_l$ and frequency $\omega$ which is physically significant for elastic waves:
        $(a)$ $\omega/\omega_0$ vs $k_l/k_0$ plots for various values of $R$; $(b)$ $\omega/\omega_0$ vs $k_l/k_0$ plots
        for $R=0.5$ predicted by Eq. (\ref{SERE53011ex}) and its atomistic counterpart.
        }%
     \label{Har1DNNN}
\end{figure}

Fig. \ref{Har1DNN} illustrates the comparison of dispersion curves predicted by atomistic theory, new NRR theory, and classical rod (CR) theory for the harmonic lattice with nearest neighbor interactions. In Fig. \ref{KnvsWn}, the wavenumber $k_l$ is plotted as a function of real valued frequency $\omega$. The wavenumber $k_l$ is generally complex in this case. Therefore, in Fig. \ref{KnvsWn} we have plotted $(\operatorname{Re}(k_l)-\operatorname{Im}(k_l))/k_0$ vs $\frac{\omega}{\omega_0}$, where $\omega_0 = \sqrt{\frac{s_a}{m_a}}$ and $k_0 = \pi/l_a$. Fig. \ref{CgvsWn} demonstrates the normalized group speed $C_g/C_0$ vs normalized frequency $\frac{\omega}{\omega_0}$ plots obtained using different theories. These plots show that both the predictions of wavenumber and group speed by new NRR theory match exactly with the same predictions (Figs. \ref{KnvsWn}-\ref{CgvsWn}) of atomistic theory for all values of wavenumbers and/or frequencies. However, the predictions by the classical rod theory fail to match with the actual predictions of atomistic theory. Results of Fig. \ref{Har1DNN} show that the new NRR equation can exactly describe the high frequency dynamics of a discrete 1D harmonic lattice with nearest neighbor interaction.
\begin{figure}[!h]
\centering
\includegraphics[scale=0.45]{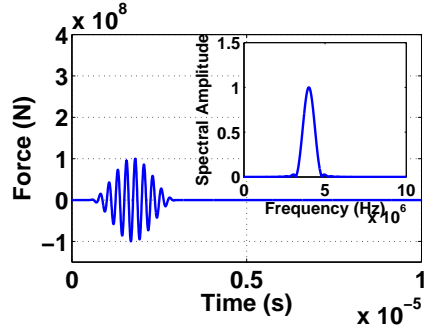}
\caption{The modulated sinusoidal pulse used in 1D harmonic lattice experiment.}
\label{mdlsinpulse}
\end{figure}
In Fig. \ref{Har1DNNN}, the effect of stiffness ratio $R_2 = s_{a2}/s_{a}$ on the overall dispersive characteristic of the 1D harmonic lattice with next-nearest-neighbor interactions is presented in detail. It can be seen in Fig. \ref{KnvsWnNNN} that the dispersive characteristic is greatly affected by the values of stiffness ratio $R_2$. This shows that the overall dispersive characteristic of an atomistic system is not only dependent on the discreteness of the system but also on the range of interactions exists among the non-nearest neighbors. Therefore, the overall dispersive nature of a material system is characterized by the prevailing heterogeneity and nonlocality in the system. Dispersion prediction by continuum equation (\ref{SERE53011ex}) is plotted and compared with the actual atomistic prediction in Fig. \ref{CgvsWnNNN}. Result in Fig. \ref{CgvsWnNNN} shows an exact match as the NRR equation (\ref{SERE53011ex}) is the exact equivalent of the discrete system.
\begin{figure}[!h]
    \label{fig:subfigures}
    \begin{center}
        \subfigure[]{%
            \label{CNNdisp}
            \includegraphics[width=0.4\textwidth]{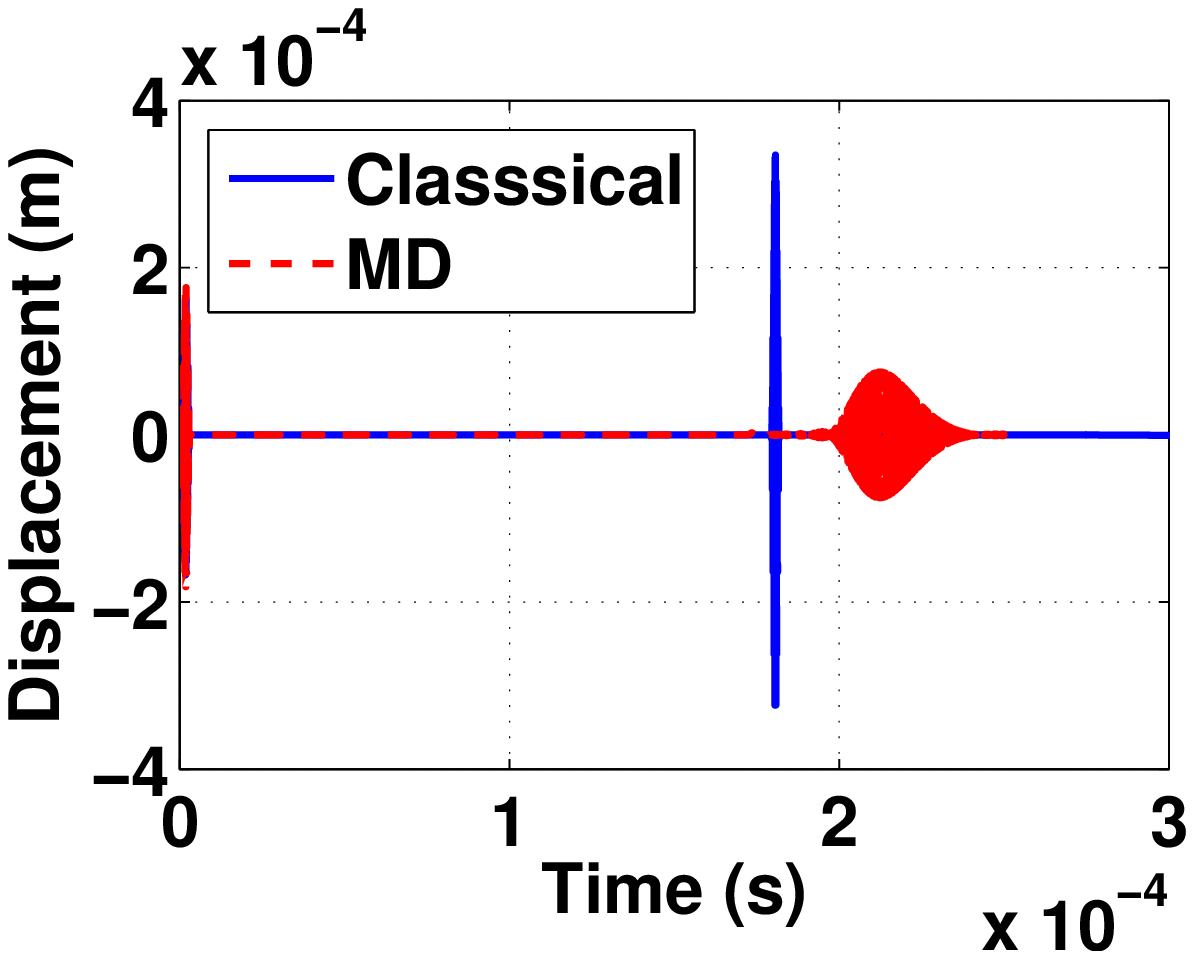}
        }%
        \subfigure[]{%
           \label{CNNvelo}
           \includegraphics[width=0.4\textwidth]{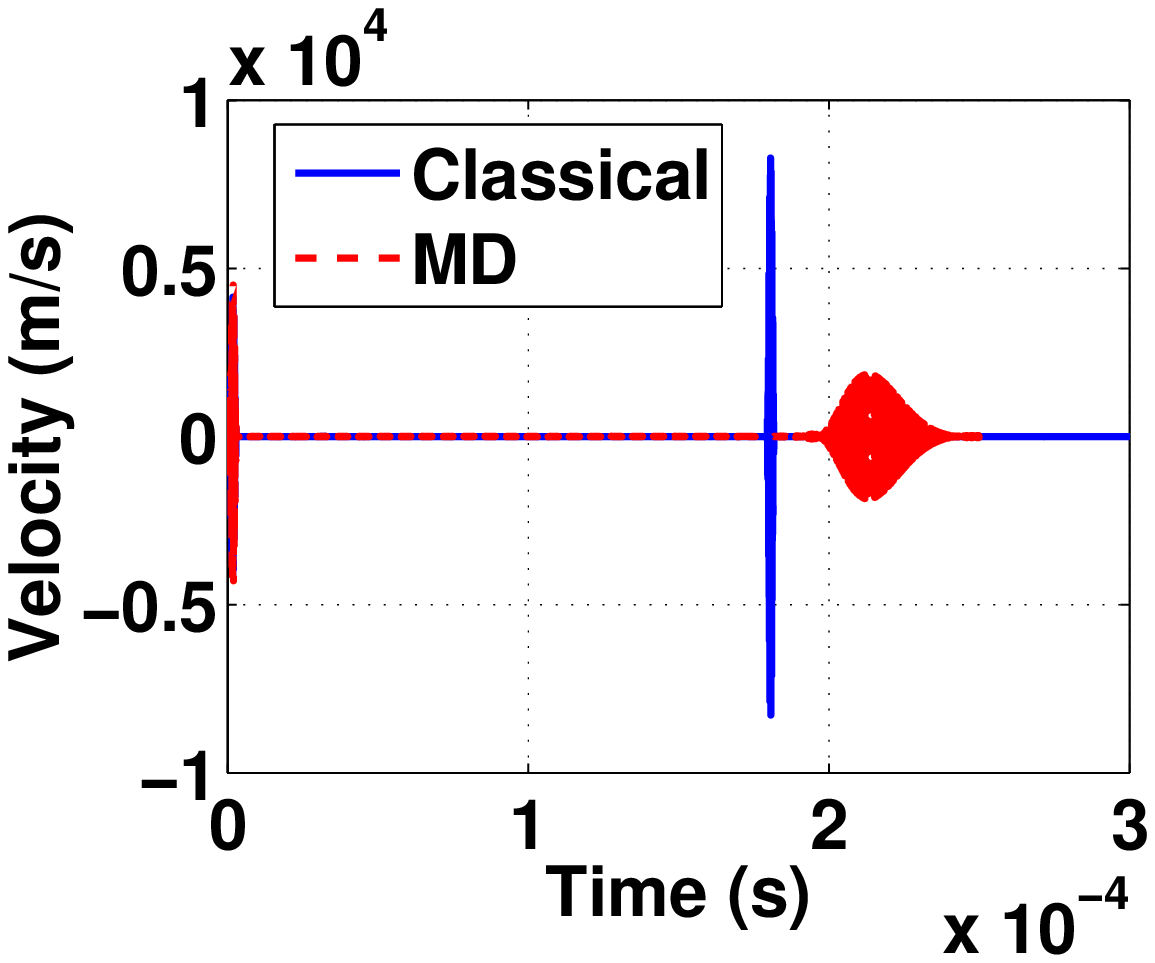}
        }
        \\ 
        \subfigure[]{%
            \label{ENNdisp}
            \includegraphics[width=0.4\textwidth]{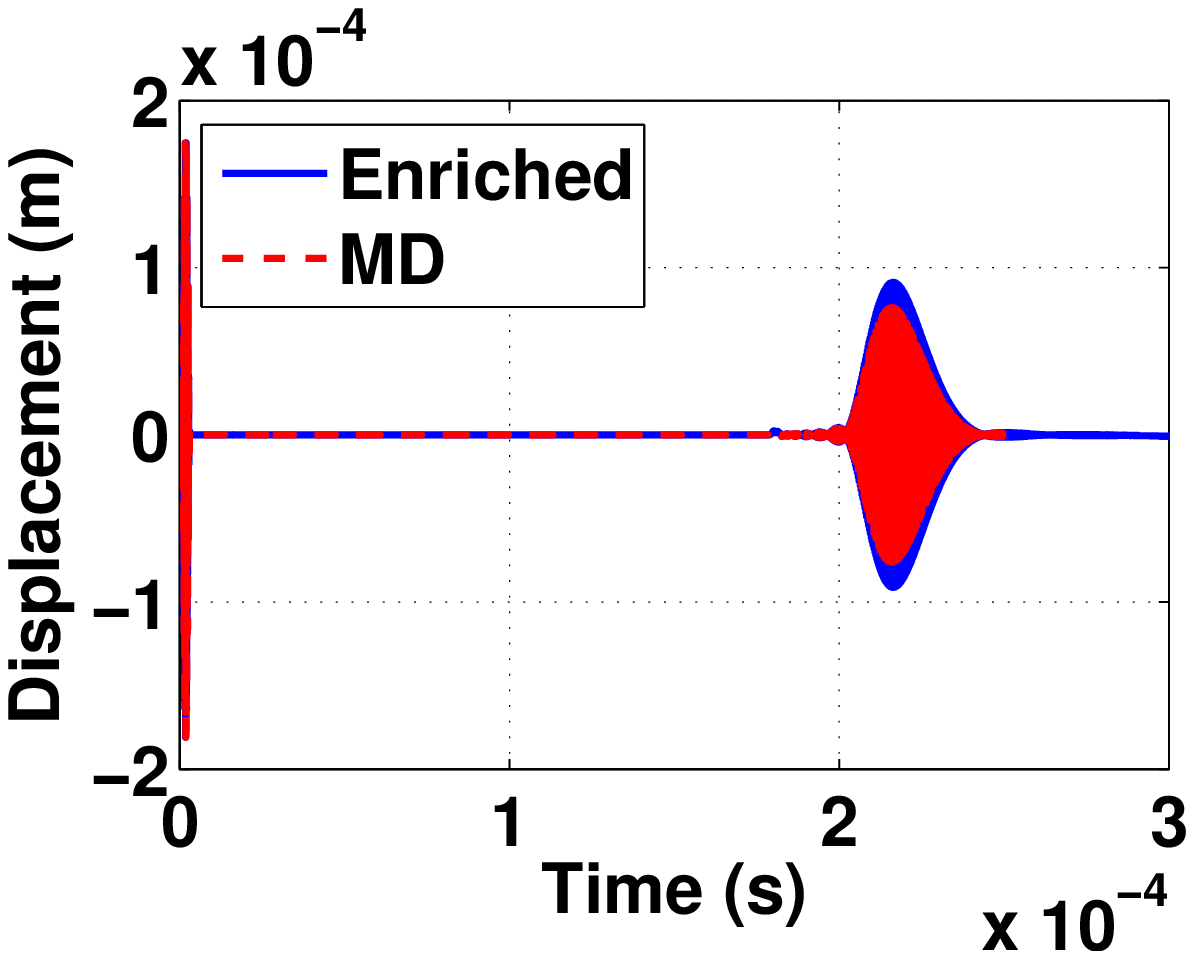}
        }%
        \subfigure[]{%
            \label{ENNvelo}
            \includegraphics[width=0.4\textwidth]{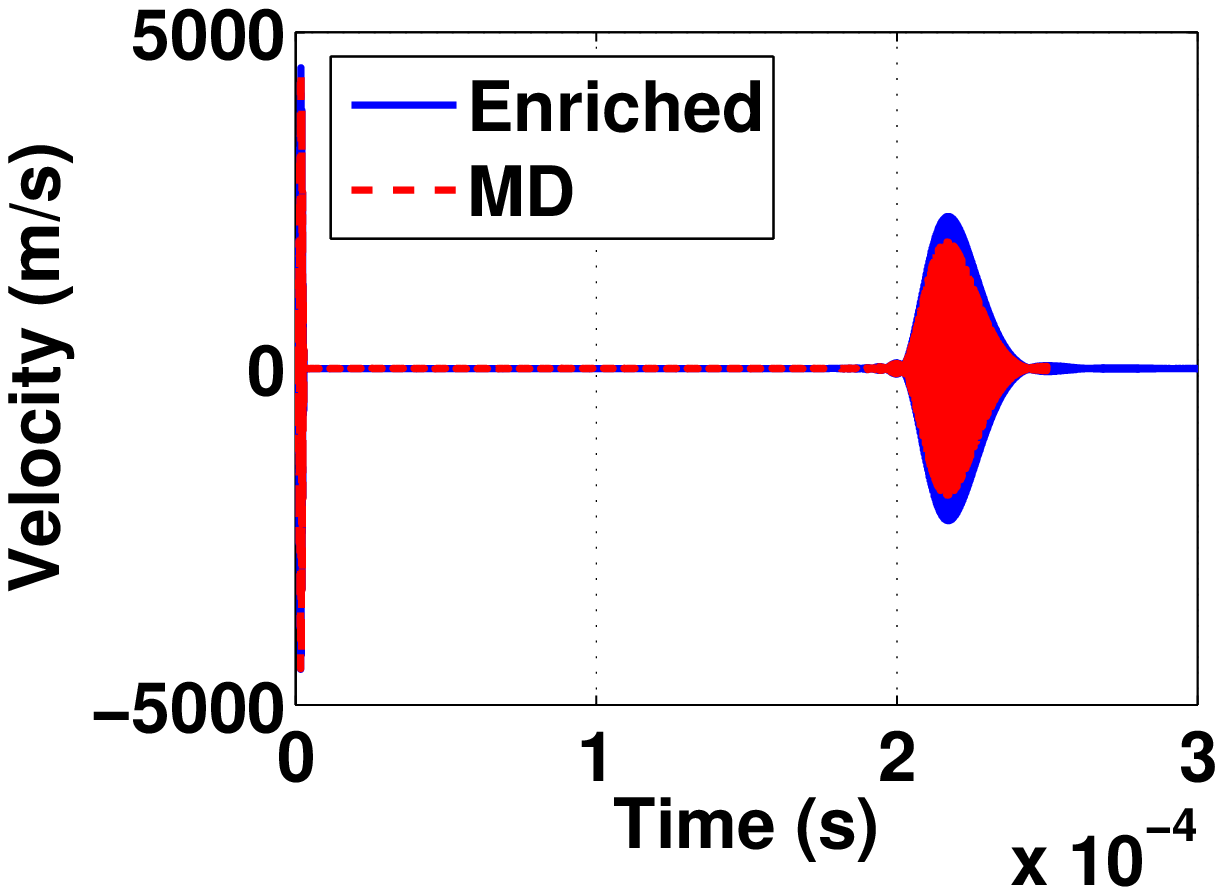}
        }%
    \end{center}
    \caption{%
        Comparison of responses of the free end atom of the $1$D lattice with nearest neighbor interaction computed by full MD and NLSFEM simulations:
        $(a)$ $u(0.5,t)$ by CR and MD; $(b)$ $\dot{u}(0.5,t)$ by CR and MD;
        $(c)$ $u(0.5,t)$ by enriched NRR and MD; $(d)$ $\dot{u}(0.5,t)$ by enriched NRR and MD.
       }%
     \label{ENN}
\end{figure}

For studying the dynamic behavior of the lattice, a modulated sinusoidal pulse (Fig. \ref{mdlsinpulse}) is applied on the rightmost atom and the displacement and velocity responses of the same atom location are measured. The pulse is modulated at the central frequency of $4\times 10^{6}$ Hz (Fig. \ref{mdlsinpulse}) which is high as compared to the $f_0=\omega_0/{2\pi}= 3.5588\times 10^{6}$ Hz. The atomistic computation is done using MD simulation which provides the standard results for this experiment. MD simulation is run for $N=10^5$ time steps with a time step length of $\Delta t = 2.5\times 10^{-9}$ s. For case-1 experiment, Eq. (\ref{SERE508ex}) is solved in NLSFEM framework using a single element for the whole lattice. The NLSFEM simulation is run for total $N= 131072$ sampling points considering $\Delta t=2.5\times 10^{-9}$ s.

Results are presented in Fig. \ref{ENN}. In Figs. \ref{CNNdisp}-\ref{CNNvelo}, displacement and velocity responses of the free end atom computed by classical rod (CR) equation and MD are presented. A significant difference in response magnitudes and propagation speeds is observed in these figures. Therefore, this comparison clearly shows that the solutions of CR equation significantly differ from the actual atomistic solutions by MD. However, solutions of the enriched NRR equation (Figs. \ref{ENNdisp}-\ref{ENNvelo}) match very satisfactorily with the atomistic solutions obtained by MD. Only a very small difference in magnitude is observed because of the comparatively large time step $\Delta t$ used in the computations. Here, the main source of error in MD simulation is the time integration scheme, which accumulates errors in computations when the time step length $\Delta t$ is not very small enough as per the requirement at very high frequency. For the selection of very small time step $\Delta t$, this negligible mismatch vanishes completely albeit with an additional cost. These results show the excellent capabilities of the new nonlocal rational continuum (NRC) equations for wave propagation problems in 1D discrete systems.
\begin{figure}[!h]
    \label{fig:subfigures}
    \begin{center}
        \subfigure[]{%
            \label{nmhr11}
            \includegraphics[width=0.45\textwidth]{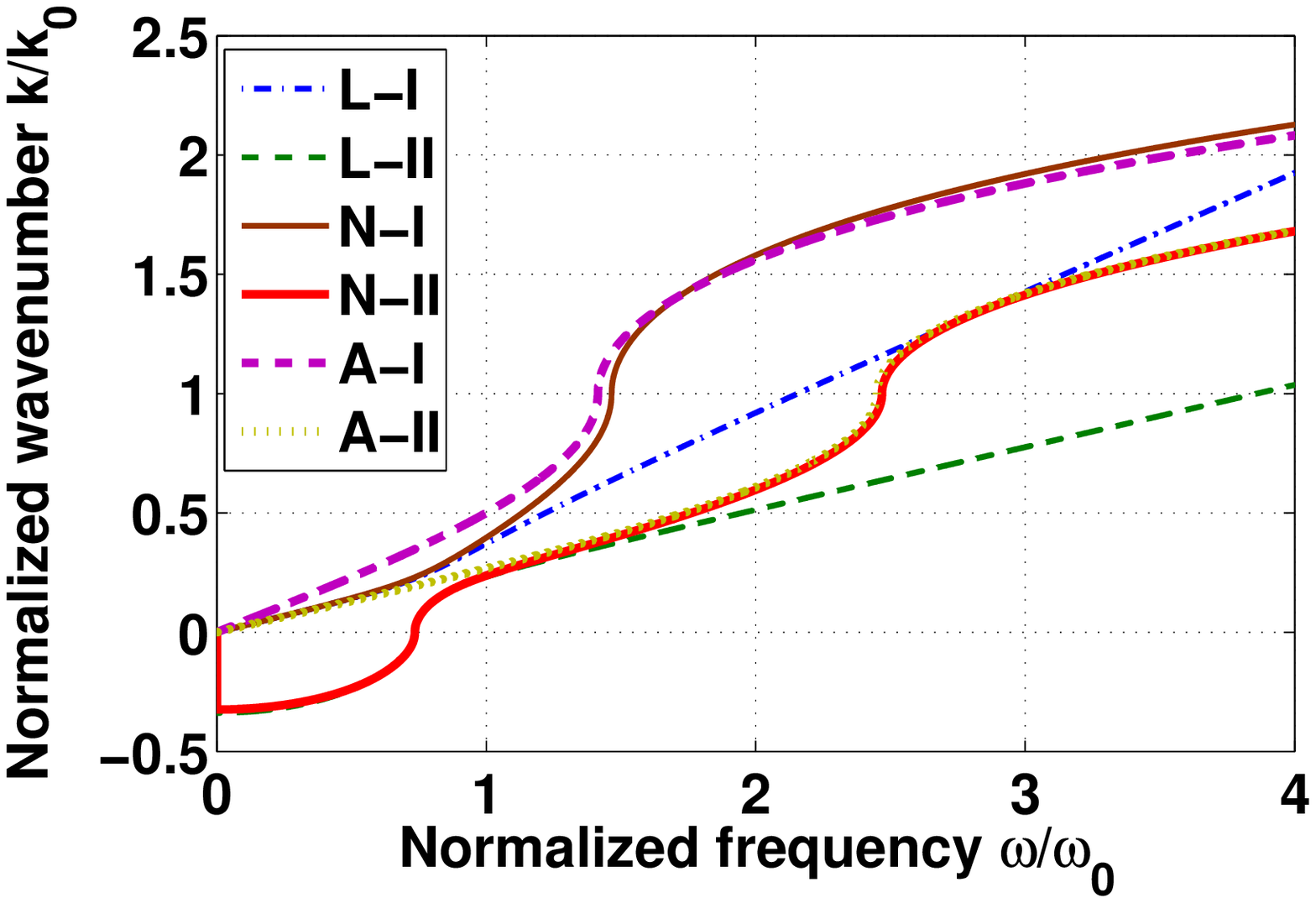}
        }%
        \subfigure[]{%
           \label{nmhr12}
           \includegraphics[width=0.45\textwidth]{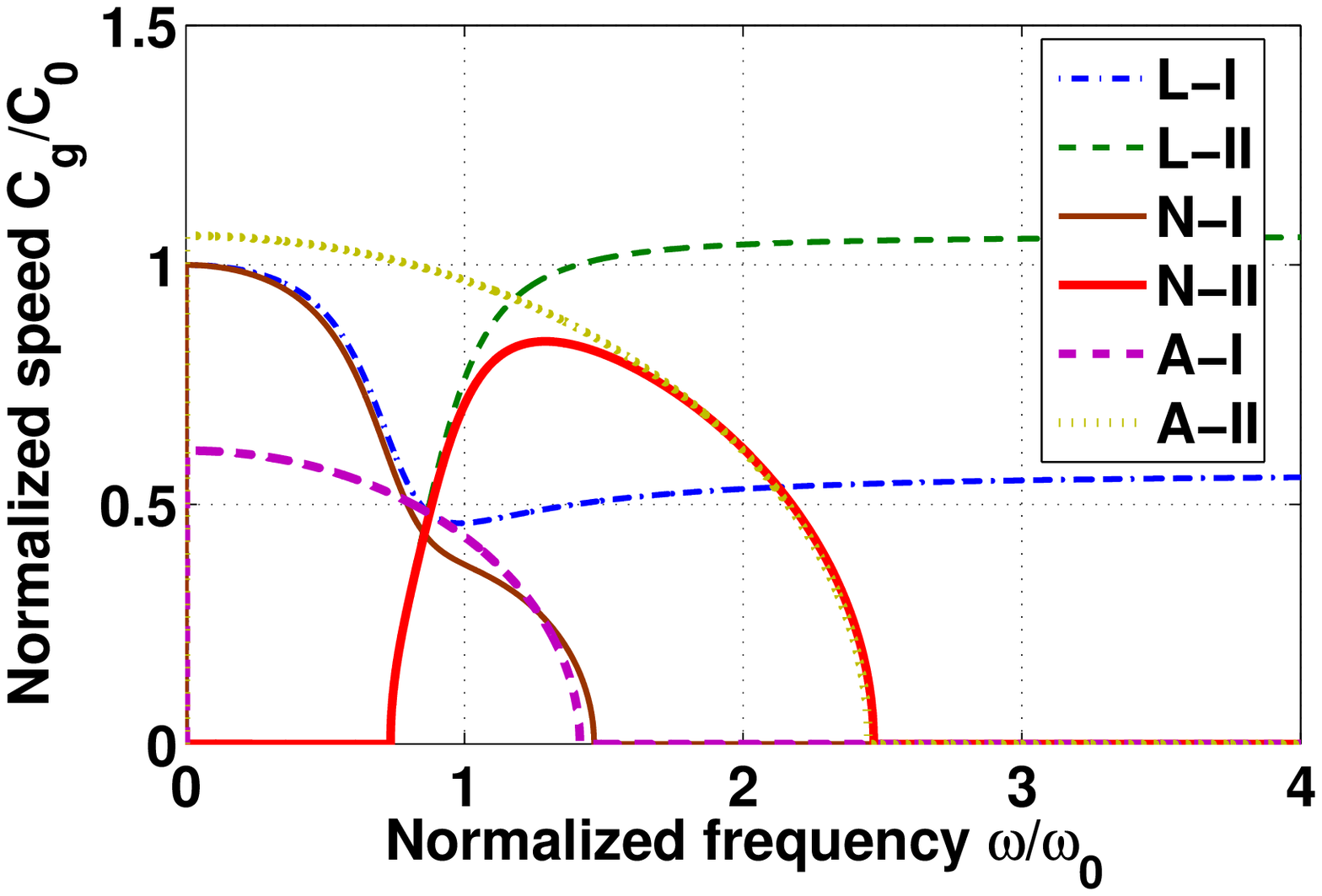}
        }
        \\ 
        \subfigure[]{%
            \label{nmhr13}
            \includegraphics[width=0.45\textwidth]{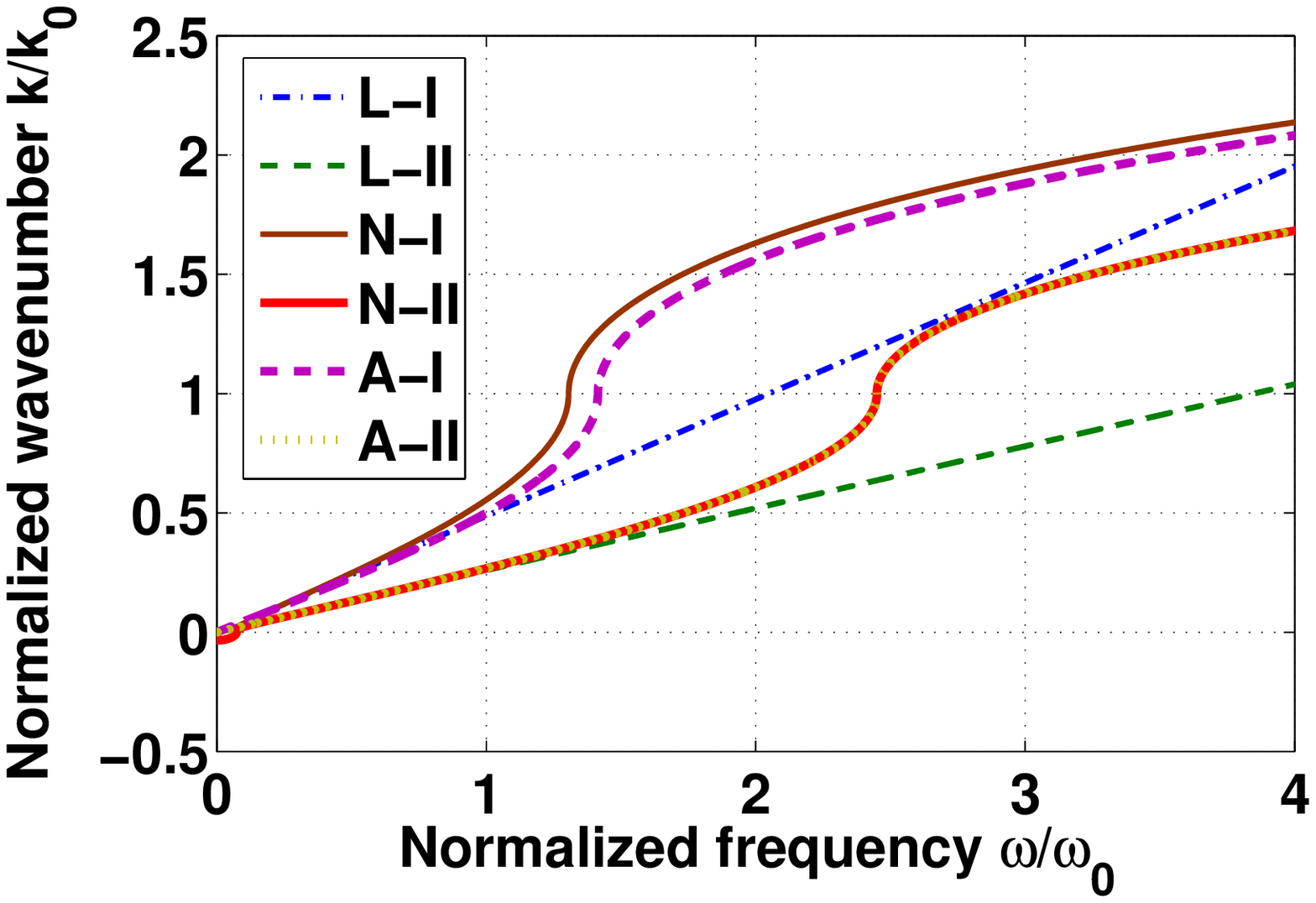}
        }%
        \subfigure[]{%
            \label{nmhr14}
            \includegraphics[width=0.45\textwidth]{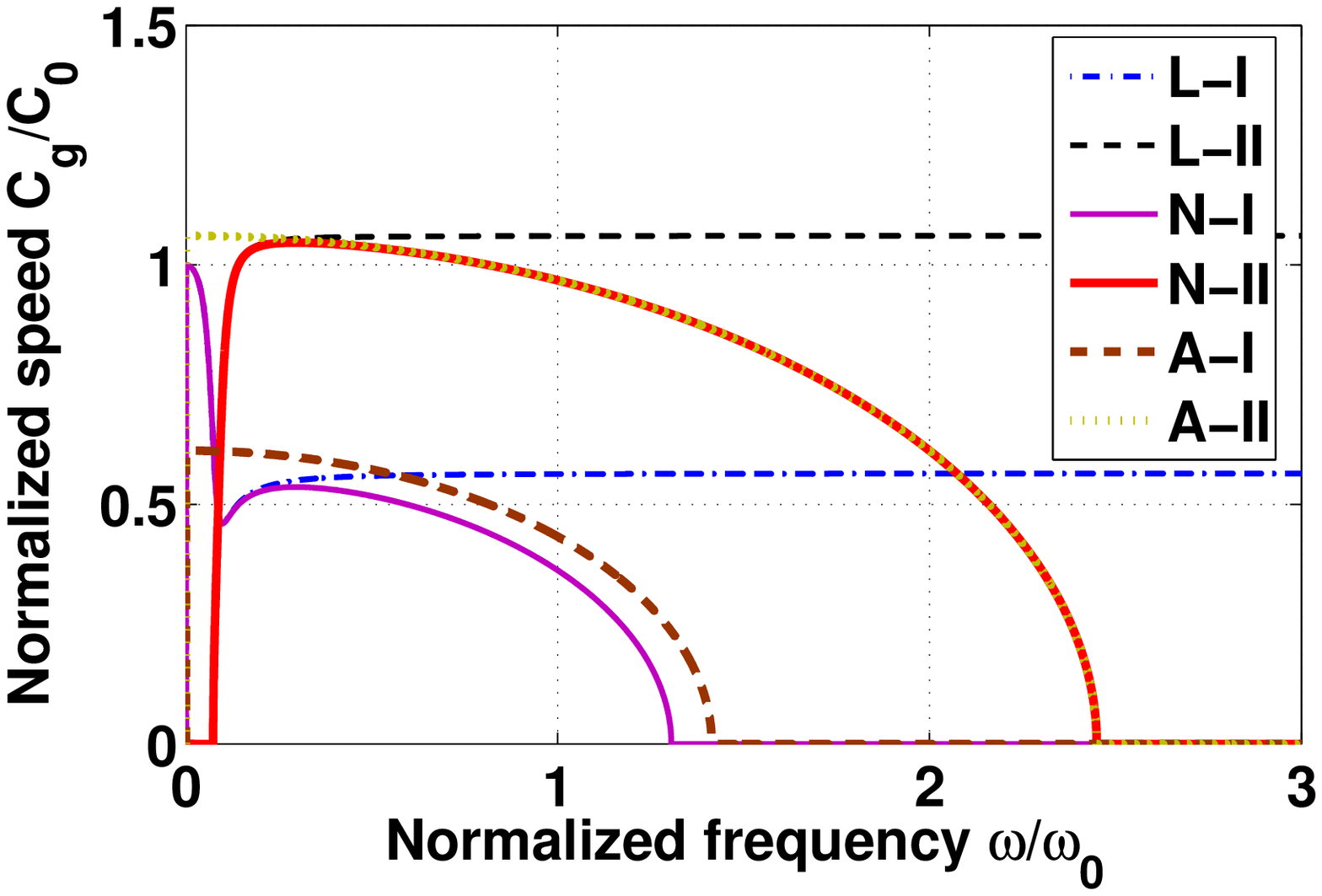}
        }%
    \end{center}
    \caption{%
        Comparison of dispersion plots predicted by the LMHR theory (denoted by L-I, and L-II), the NRMHR theory (denoted by N-I, and N-II), and the atomistic theory (denoted by A-I, and A-II) for the 2D atomistic lamina with different $r_0/d$ ratios. Here, I, and II represent the first and second branches of the dispersion predictions, respectively:
        $(a)$ $(Re(k)-Im(k))/k_0$ vs $\omega/\omega_0$ for $r_0/d=0.2$; $(b)$ $C_g/C_0$ vs $\omega/\omega_0$ for $r_0/d=0.2$; $(c)$ $(Re(k)-Im(k))/k_0$ vs $\omega/\omega_0$ for $r_0/d=0.02$; and $(d)$ $C_g/C_0$ vs $\omega/\omega_0$ for $r_0/d=0.02$.
        Here, $k_0 = \pi/r_x$, $C_0=r_0\omega_0$, and $\omega_0 = \sqrt{S_1/m_a}$.
        }%
     \label{EMHR1}
\end{figure}

\subsection{Enhanced dispersion relation of the nonlocal rational Mindlin-Herrmann rod model}
\label{NRMHRanalysis}
For studying the enhanced discrete dispersion characteristics of the newly proposed nonlocal rational Mindlin-Herrmann rod (NRMHR) model, a $2$D triangular lattice system with hexagonal nearest neighbor interaction and with an equilibrium atomic separation $r_0 = 2 \times 10^{-10}$ m is considered. The $2$D lamina consists of atoms of mass $m_a = 1 \times 10^{-26}$ kg lying in total $N_c\geq4$ even number of columns parallel to the $Y$-directions (Fig \ref{2Dhcp}). Every odd and even numbered column has $(n+1)$ and $(n+2)$ atoms, respectively. Here, $n\geq1$ is any integer number. The $2$D lamina has a length of $L=L_x= (N_c-1)\frac{\sqrt{3}}{2}r_0$ m, a breadth of $d=L_y= (n+2)r_0$ m and a thickness $h = 1\times 10^{-10}$ m is assumed. The $2$D atomistic system is idealized as a deep axial waveguide under loading along the $X$-direction, where the wave is propagating in the X-direction only \citep{patra2014spectral}. It is reasonable  to consider a harmonic potential $U_1$ of spring constant $ S_1 = 5$ N/m for the small strain consideration. Considering a plane stress condition, the values of the elastic constants are computed as $E=\frac {2}{\sqrt{3}} \frac {S_1}{h}$, $G=\frac {\sqrt{3}}{4} \frac {S_1}{h}$, and Poisson's ratio $\nu = \frac{1}{3}$. The adjustable parameters for the present system are assumed as $\gamma_1 = 0.7$ and $\gamma_2 = 1$.
\begin{figure}[!h]
    \label{fig:subfigures}
    \begin{center}
        \subfigure[]{%
            \label{nmhr21}
            \includegraphics[width=0.45\textwidth]{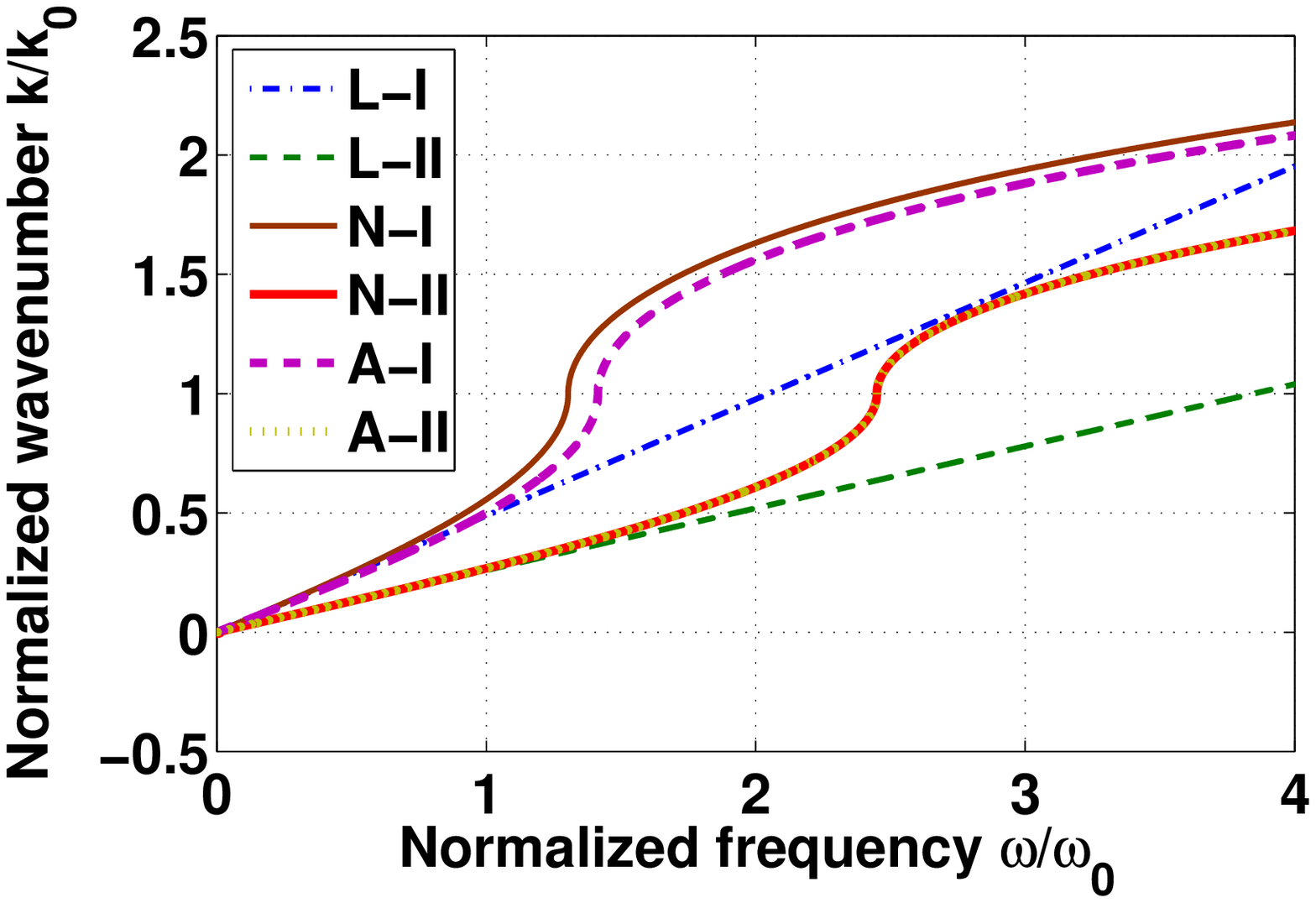}
        }%
        \subfigure[]{%
           \label{nmhr22}
           \includegraphics[width=0.45\textwidth]{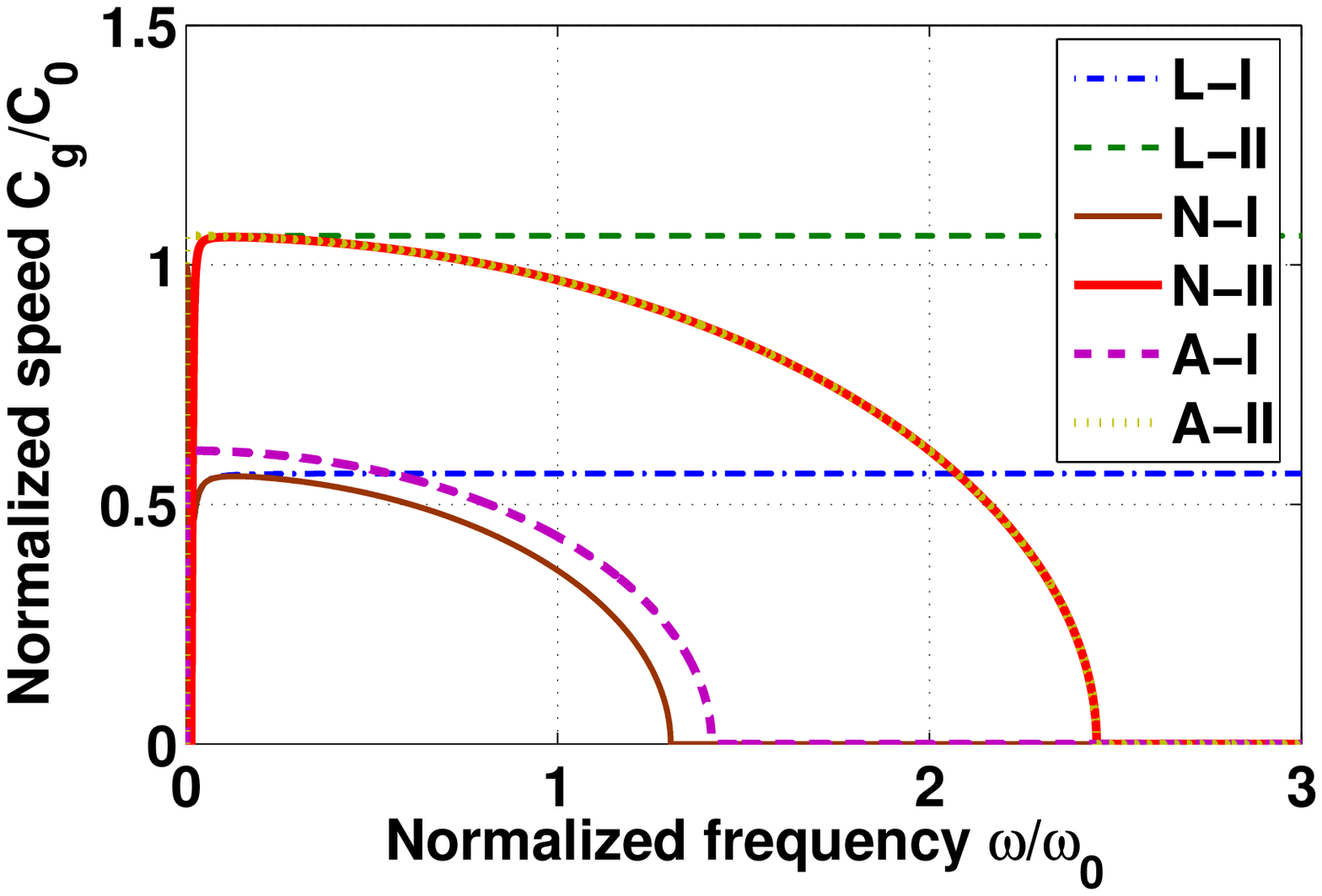}
        }
        \\ 
%
    \end{center}
    \caption{%
        Comparison of dispersion curves predicted by three different theories for modeling the axial wave propagation in a 2D atomistic lamina with $r_0/d= 0.002$:
        $(a)$ $(Re(k)-Im(k))/k_0$ vs $\omega/\omega_0$ and $(b)$ $C_g/C_0$ vs $\omega/\omega_0$.
        }%
     \label{EMHR2}
\end{figure}

In the present study, only nearest neighbor interaction is considered for comparing the dispersion relation of the NRMHR equations with the atomistic dispersion relation of a 2D lamina. The dispersion relations for NRMHR model are plotted using Eqs. (\ref{SEBE5036})-(\ref{SEBE5037}). The atomistic counterparts of the same are plotted using Eqs. (\ref{SEBE50310})-(\ref{SEBE50311}). Figs. \ref{EMHR1}-\ref{EMHR2} illustrate the comparison of dispersion relations obtained using different theories for varying $r_0/d$ ratios. Figs. \ref{nmhr11}-\ref{nmhr14} show that the dispersion predictions of NRMHR equation slightly differ from the predictions of classical Mindlin-Herrmann (LMHR) theory before the cut-on frequencies. However, they differ drastically from each other after the cut-on frequencies. Furthermore, the predictions of the new NRMHR theory agree very well with the predictions of atomistic theory after the cut-on frequencies (Figs. \ref{nmhr11} - \ref{nmhr22}). It can be seen from Figs. \ref{nmhr11}-\ref{nmhr12} that the cut-on frequency becomes very high ($f_{cut-on}=2.615\times 10^{12}$ Hz) for very slender rods (i.e., for $d\leq5\times r_0$) and the predictions of NRMHR theory agree very well with the atomistic predictions at higher frequencies. However, for a very slender rod, the first branch predictions of NRMHR theory show better agreement with the atomistic prediction (Figs. \ref{nmhr11}-\ref{nmhr12}) after the cut-on frequency. For lower $r_0/d$ ratios (Figs. \ref{nmhr13}-\ref{nmhr22}), cut-on frequency values decrease rapidly and the predictions of the NRMHR model match very accurately with the atomistic predictions over the greater portion of the physically significant frequency range. Interestingly, the second mode (i.e., the contraction mode) predictions of this newly proposed NRMHR theory match very accurately with the second mode predictions of the atomistic theory after its cut-on frequencies. However, the first mode i.e., the longitudinal mode predictions of the enriched NRMHR theory are in quite good agreement with the predictions of atomistic theory over the entire range of physically significant frequency range. Moreover, it is clearly observed in Figs. \ref{nmhr12},\ref{nmhr14}, and \ref{nmhr22} that the NRMHR theory can very accurately capture the band-gap characteristics of wave dispersion. It can very efficiently model the high-frequency dynamic behavior of atomistic systems of several nanometer size. Therefore, the nonlocal rational continuum theory can be used to design novel metamaterials, and micro/nano material devices.
\begin{figure}[!h]
    \label{fig:subfigures}
    \begin{center}
        \subfigure[]{%
            \label{nrtb11}
            \includegraphics[width=0.45\textwidth]{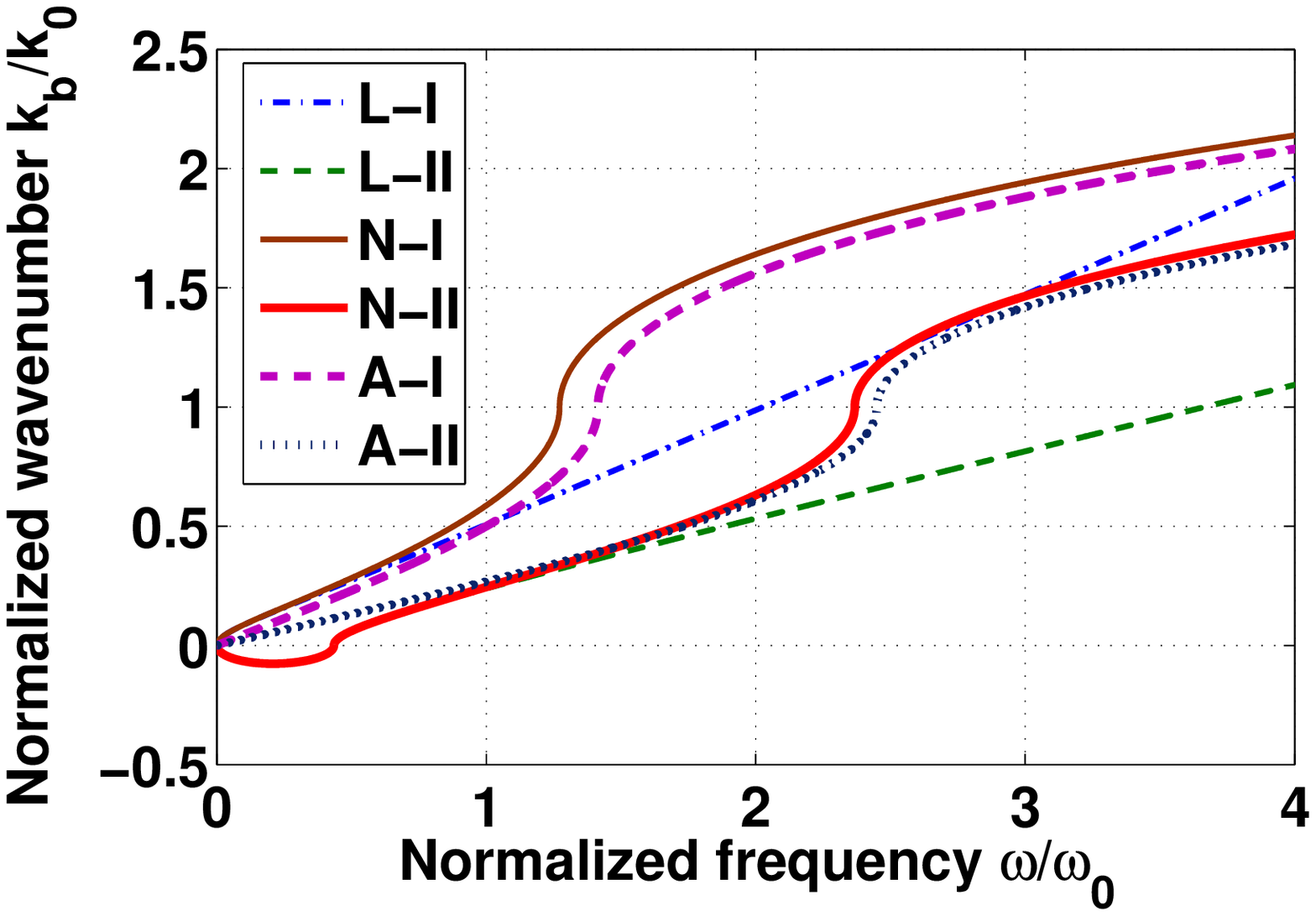}
        }%
        \subfigure[]{%
           \label{nrtb12}
           \includegraphics[width=0.45\textwidth]{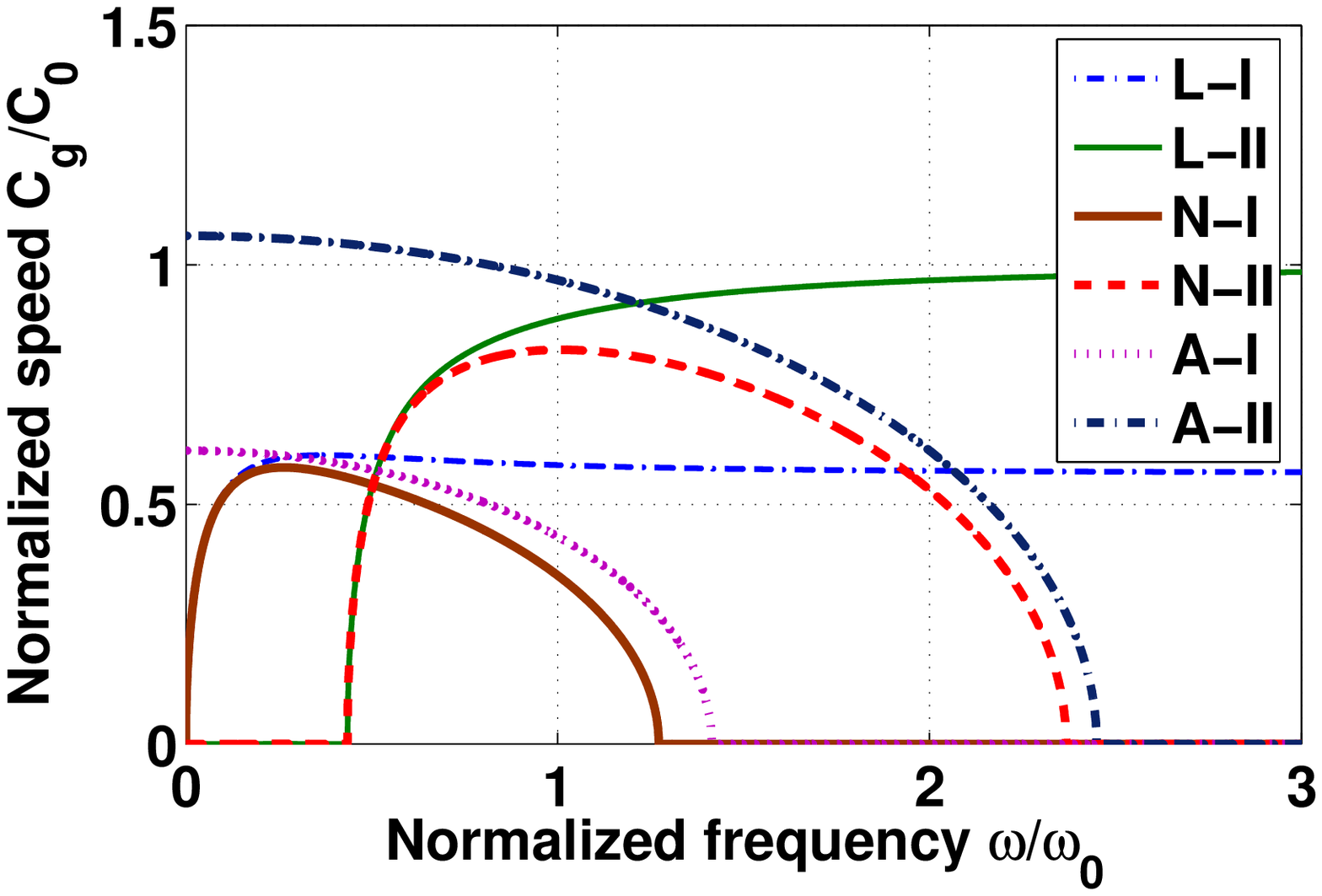}
        }
        \\ 
        \subfigure[]{%
            \label{nrtb13}
            \includegraphics[width=0.45\textwidth]{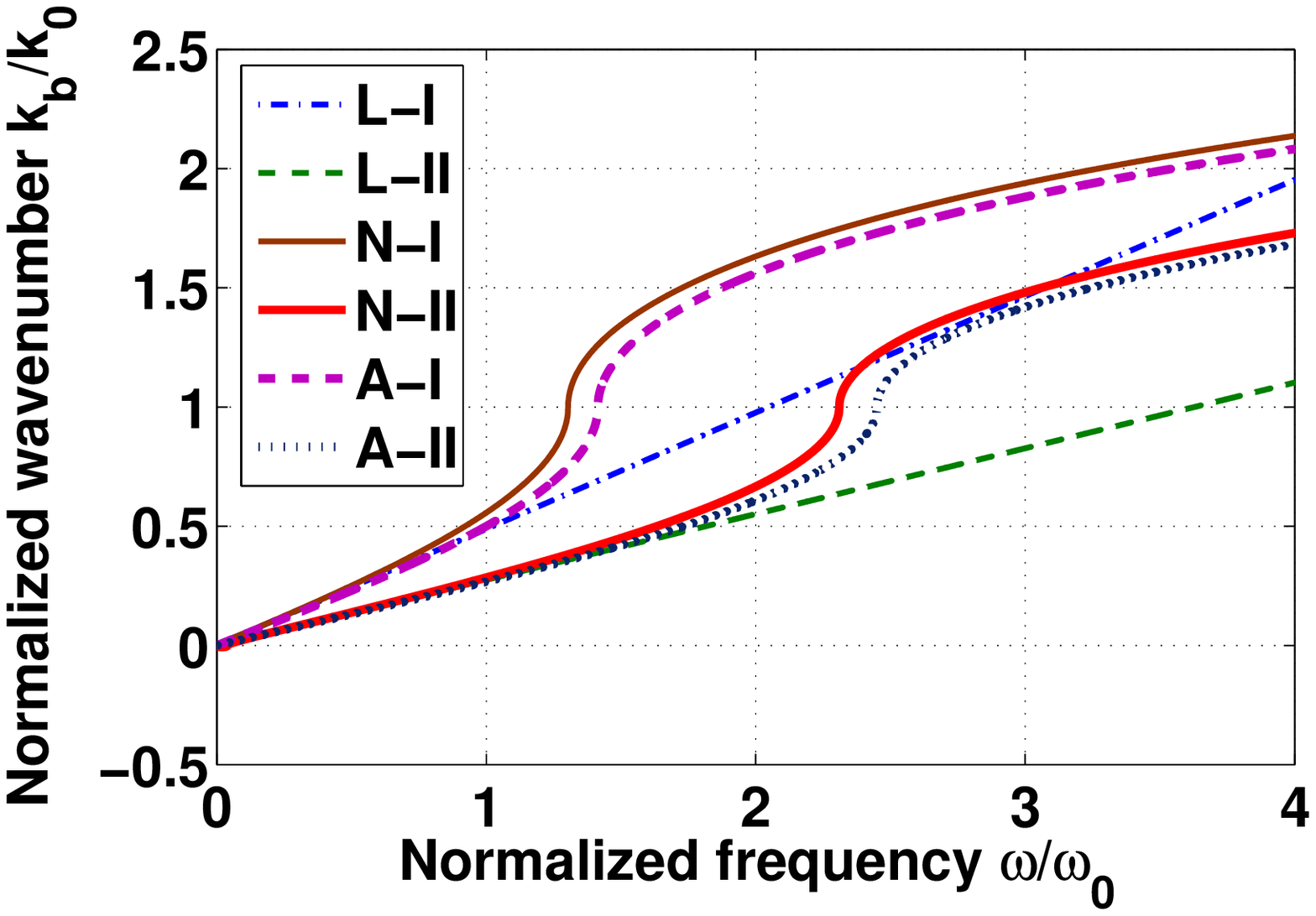}
        }%
        \subfigure[]{%
            \label{nrtb14}
            \includegraphics[width=0.45\textwidth]{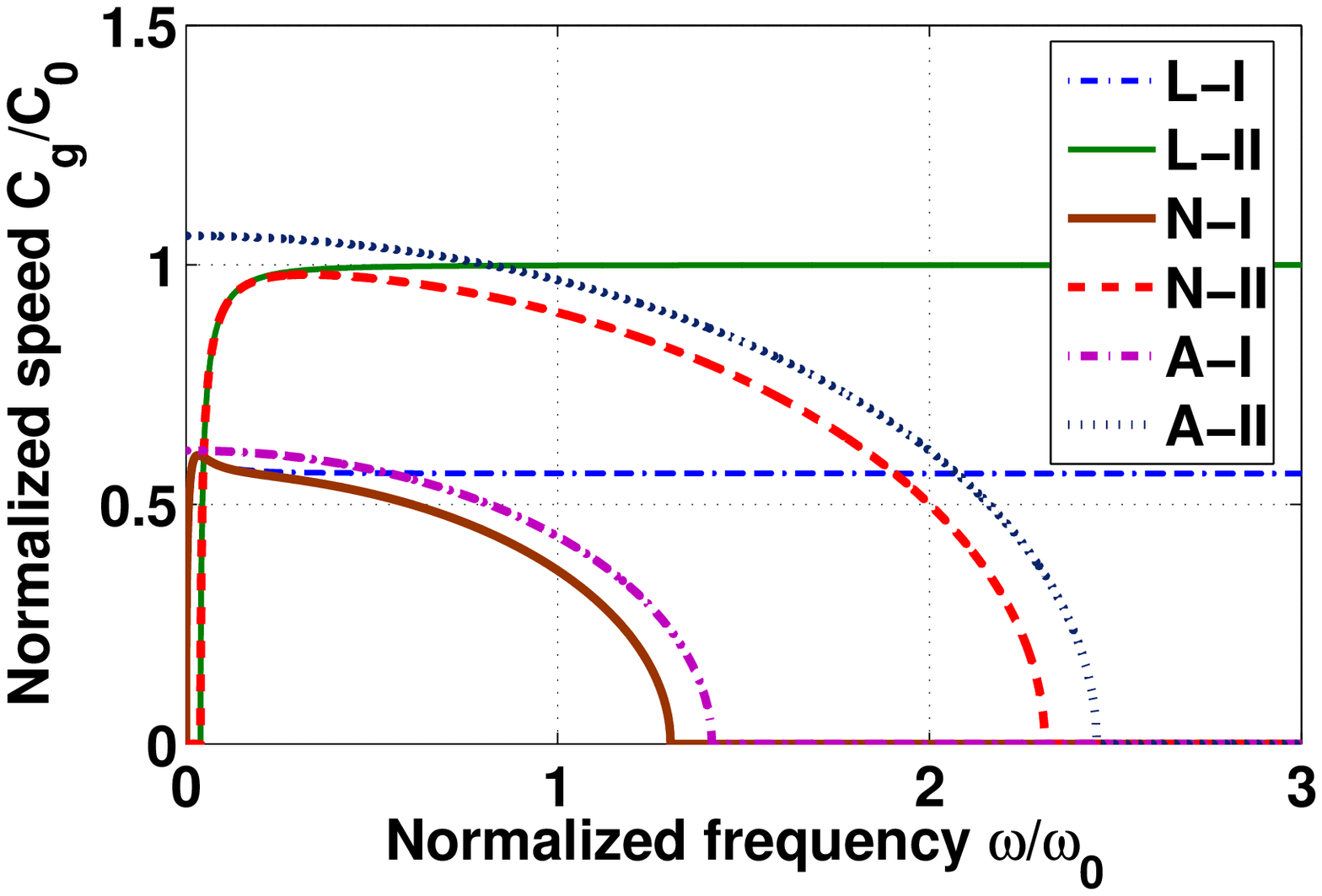}
        }%
    \end{center}
    \caption{%
        Comparison of dispersion curves predicted by the LTB theory (denoted by L-I, and L-II), the NRTB theory (denoted by N-I, and N-II), and the atomistic theory (denoted by A-I, and A-II) for the 2D atomistic lamina with different $r_0/d$ ratios. Here, I and II denote first and second branches of dispersion relationships, respectively:
        $(a)$ $(Re(k_b)-Im(k_b))/k_0$ vs $\omega/\omega_0$ for $r_0/d=0.2$;
        $(b)$  $C_g/C_0$ vs $\omega/\omega_0$ for $r_0/d=0.2$;
        $(c)$ $(Re(k_b)-Im(k_b))/k_0$ vs $\omega/\omega_0$ for $r_0/d=0.02$;
        $(d)$ $C_g/C_0$ vs $\omega/\omega_0$ for $r_0/d=0.02$;
       }%
     \label{RETBNN1}
\end{figure}

\subsection{Enhanced dispersive characteristics of the nonlocal rational Timoshenko beam model}
\label{NRTBanalysis}
The $2$D system of Section $3.2$ can be idealized as a nonlocal rational Timoshenko beam (NRTB) for a propagating wave in the $X$-direction under a transverse loading along the $Y$-direction. Similar small strain condition is assumed for this analysis. All the physical parameters and conditions for the system are kept same as in the previous study. The adjustable parameters are assumed as $K_1 = 0.85$ and $K_2 = 1$ \citep{gopalakrishnan1992matrix}.
In the present case, Eqs. (\ref{SEBE5044b})-(\ref{SEBE50410b}) describe the dynamics of the 2D atomistic lamina appropriately for transverse vibration propagation along the $X$-axis.
\begin{figure}[!h]
    \label{fig:subfigures}
    \begin{center}
        \subfigure[]{%
            \label{nrtb21}
            \includegraphics[width=0.45\textwidth]{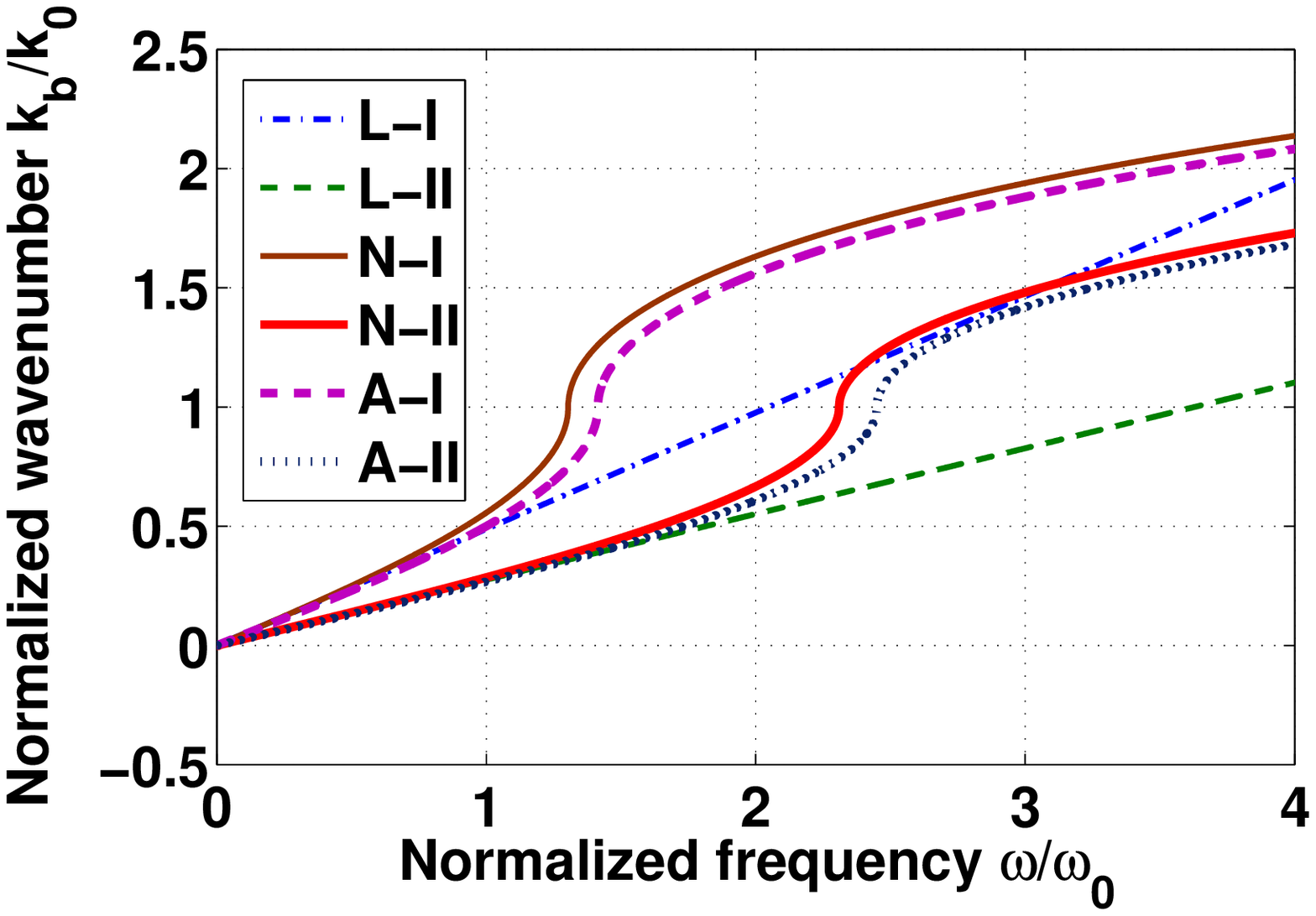}
        }%
        \subfigure[]{%
           \label{nrtb22}
           \includegraphics[width=0.45\textwidth]{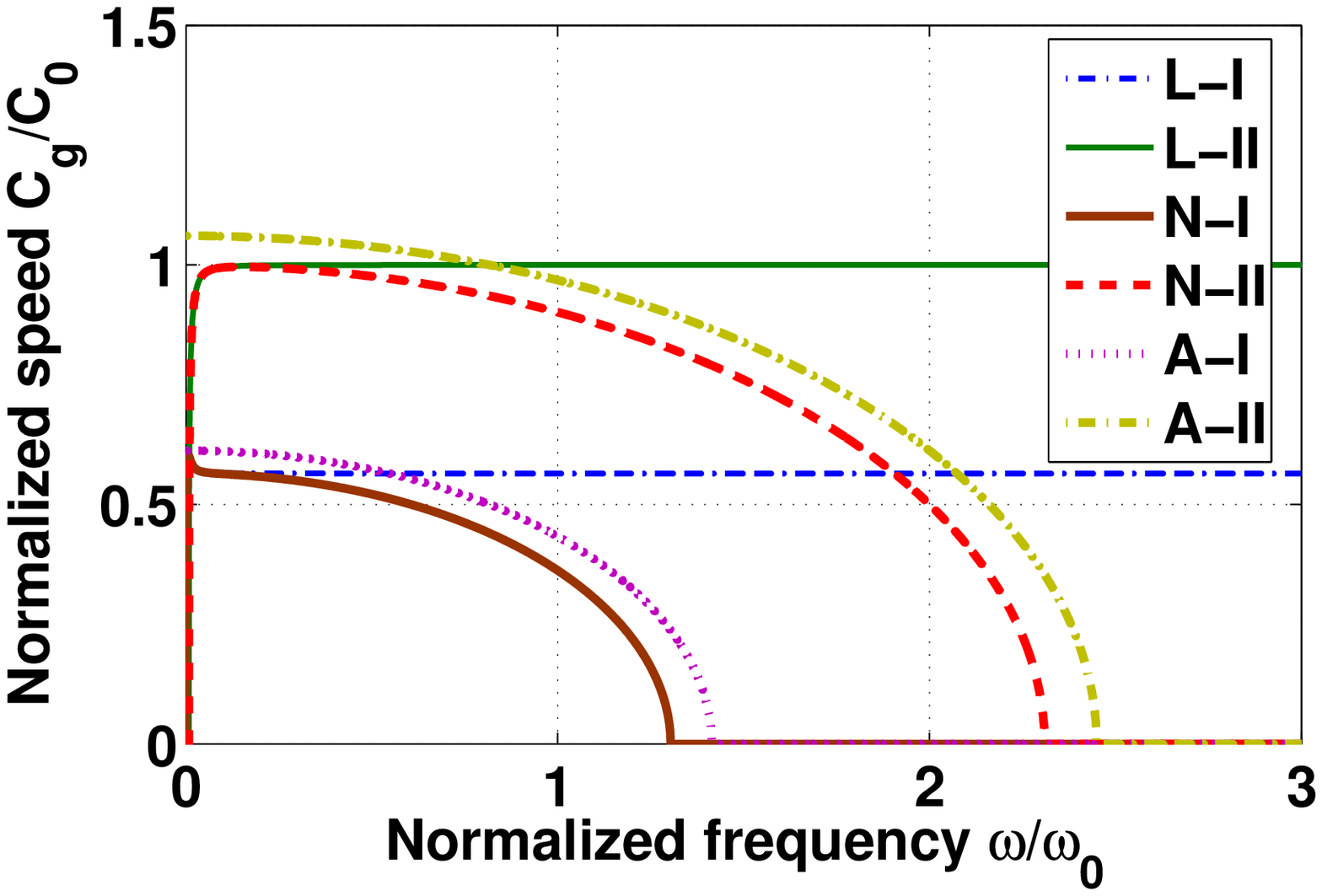}
        }
        \\ 
%
    \end{center}
    \caption{%
        Comparison of dispersion plots predicted by the atomistic equations, the LTB equations, and the NRTB equations (\ref{SEBE5049b}-\ref{SEBE50410b}) for the $2$D atomistic lamina:
        $(a)$ $(Re(k_b)-Im(k_b))/k_0$ vs $\omega/\omega_0$ for $r_0/d=0.002$;
        $(b)$ $C_g/C_0$ vs $\omega/\omega_0$ for $r_0/d=0.002$;
       }%
     \label{RETBNN2}
\end{figure}

Thus, the dispersion relations for the enriched NRTB equations are obtained using Eqs. (\ref{SEBE5049b})-(\ref{SEBE50410b}). The atomistic counterparts of the same system are obtained assuming the atomistic displacement fields of the form
\begin{align}\label{nrTB5}
u_{i,n} = -y_n\hat{\phi}_0 e^{-i_m(k_b ir_x - \omega t)} = -(j-j_0)r_y\hat{\phi}_0 e^{-i_m(k_b ir_x - \omega t)}, \;\; v_{i,n} = \hat{v}_0 e^{-i_m(k_b ir_x - \omega t)}
\end{align}
From Eqs. (\ref{SEBE5041b}) and (\ref{SEBE5026}), atomistic dispersion relationship is obtained for a beam type model of the $2$D lamina. Figs. \ref{RETBNN1}-\ref{RETBNN2} illustrate the comparison of dispersion relations obtained using different theories. It can be seen in Figs. \ref{nrtb11}-\ref{nrtb12} that for very slender beams i.e., for higher $r_0/d$ ratios the cut-on frequencies become very high ($f_c=1.55\times 10^{12}$ Hz, for $r_0/d=0.2$) and the first branch of the dispersion predictions by NRTB theory agree well with the atomistic counterpart over the entire frequency range. The second branch prediction of NRTB shows very well agreement with its atomistic counterpart only after the cut-on frequency. For a very slender beam, both the branches of dispersion predictions by NRTB show better agreement with their atomistic counterparts at higher frequencies (Figs. \ref{nrtb11}-\ref{nrtb12}). The cut-on frequency values decrease significantly with lower $r_0/d$ ratios (Figs. \ref{nrtb13}-\ref{nrtb22}), and the predictions of the NRTB theory agree well with the atomistic predictions over the greater portion of physically significant frequency band. It is clearly observable in Figs. \ref{nrtb11}-\ref{nrtb22} that the dispersion branches predicted by NRTB theory differ considerably from their counterparts predicted by classical Timoshenko beam (LTB) theory at higher frequencies. Wavenumber and group velocity predictions of the LTB theory become significantly erroneous at higher frequencies. Moreover, the classical Timoshenko beam theory cannot describe the maximum feasible frequency bound of wave propagation which is the reality for an actual atomistic system. Although with a small deviation, the wavenumber and group velocity predictions of the NRTB follow the same trend as the atomistic predictions (Figs. \ref{nrtb11}-\ref{nrtb22}) beyond the cut-on frequency of the system. Interestingly, the dominant mode (i.e., the shear mode) predictions of the enriched NRTB show good agreement with the shear mode predictions of the atomistic theory over a greater portion of the entire physically significant frequency range. Moreover, the NRTB theory gives the maximum feasible frequency bounds for its different wave modes with small errors as compared to its atomistic counterparts. Therefore, these results suggest the physically significant applicability of the newly proposed NRTB in modeling high frequency dynamics of many micro/nano-mechanical systems as well as in modeling of many macro systems.
\begin{figure}[!h]
    \label{fig:subfigures}
    \begin{center}
        \subfigure[]{%
            \label{ssb1}
            \includegraphics[width=0.45\textwidth]{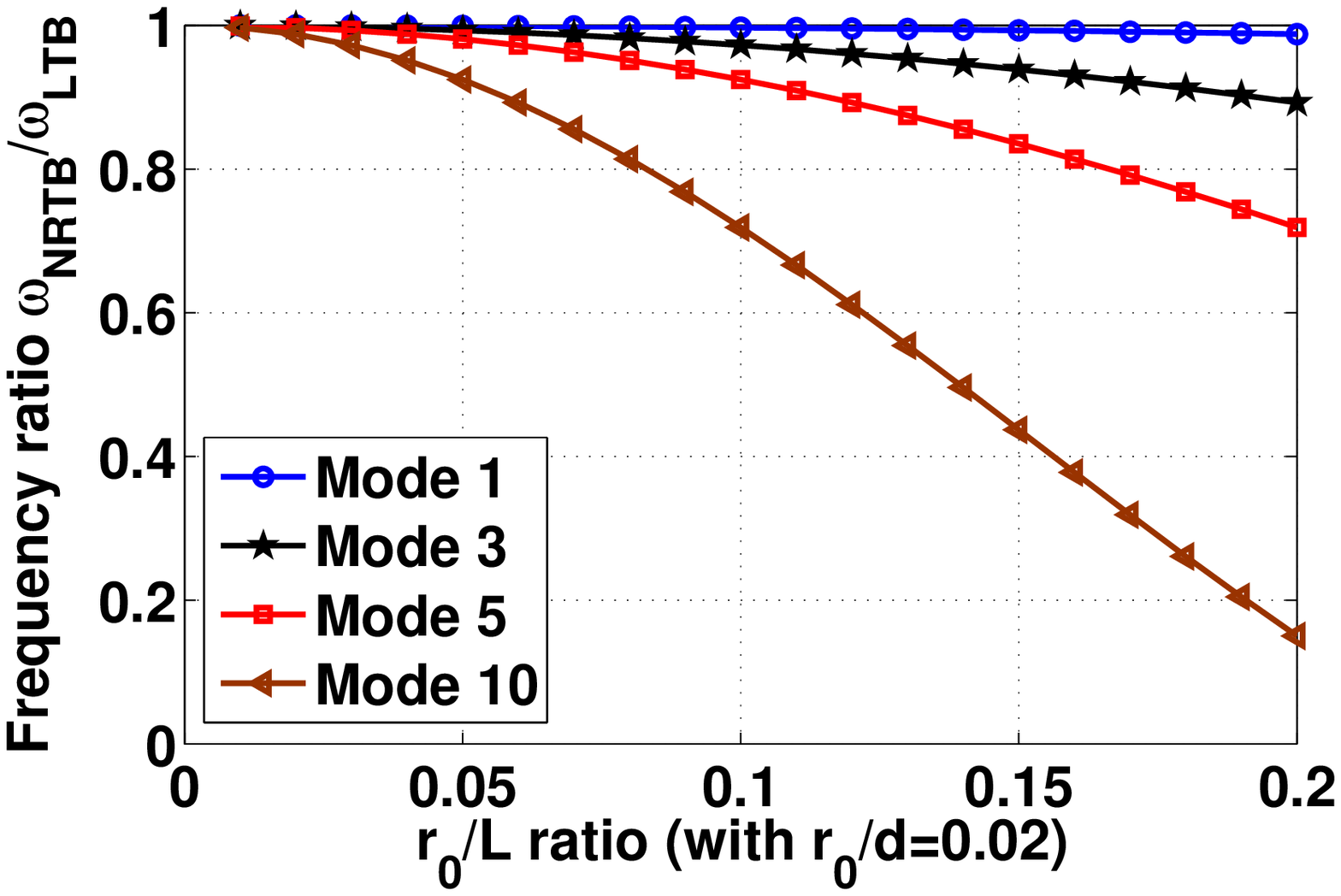}
        }%
        \subfigure[]{%
           \label{ssb2}
           \includegraphics[width=0.45\textwidth]{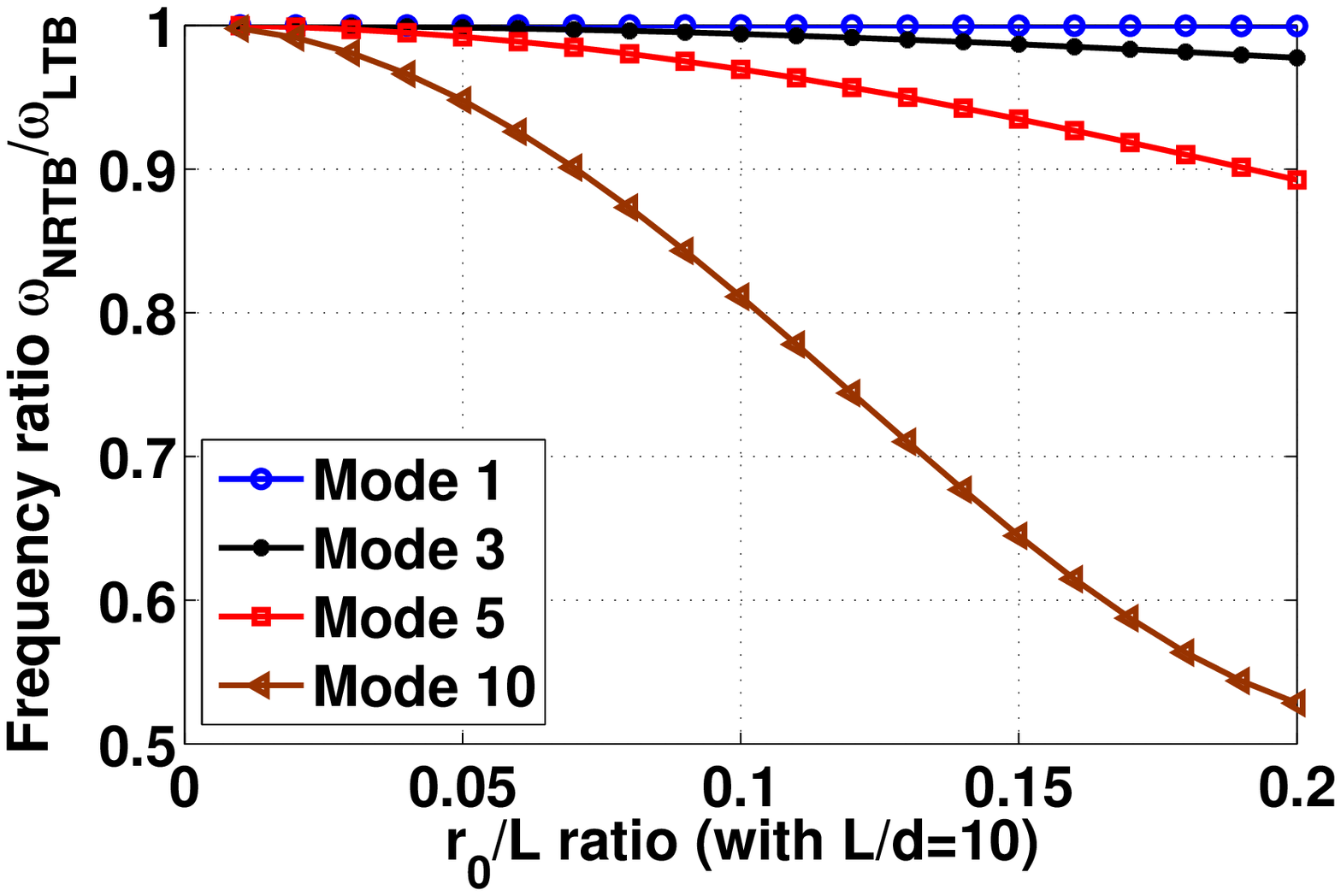}
        }
        \\ 
        \subfigure[]{%
            \label{ssb3}
            \includegraphics[width=0.45\textwidth]{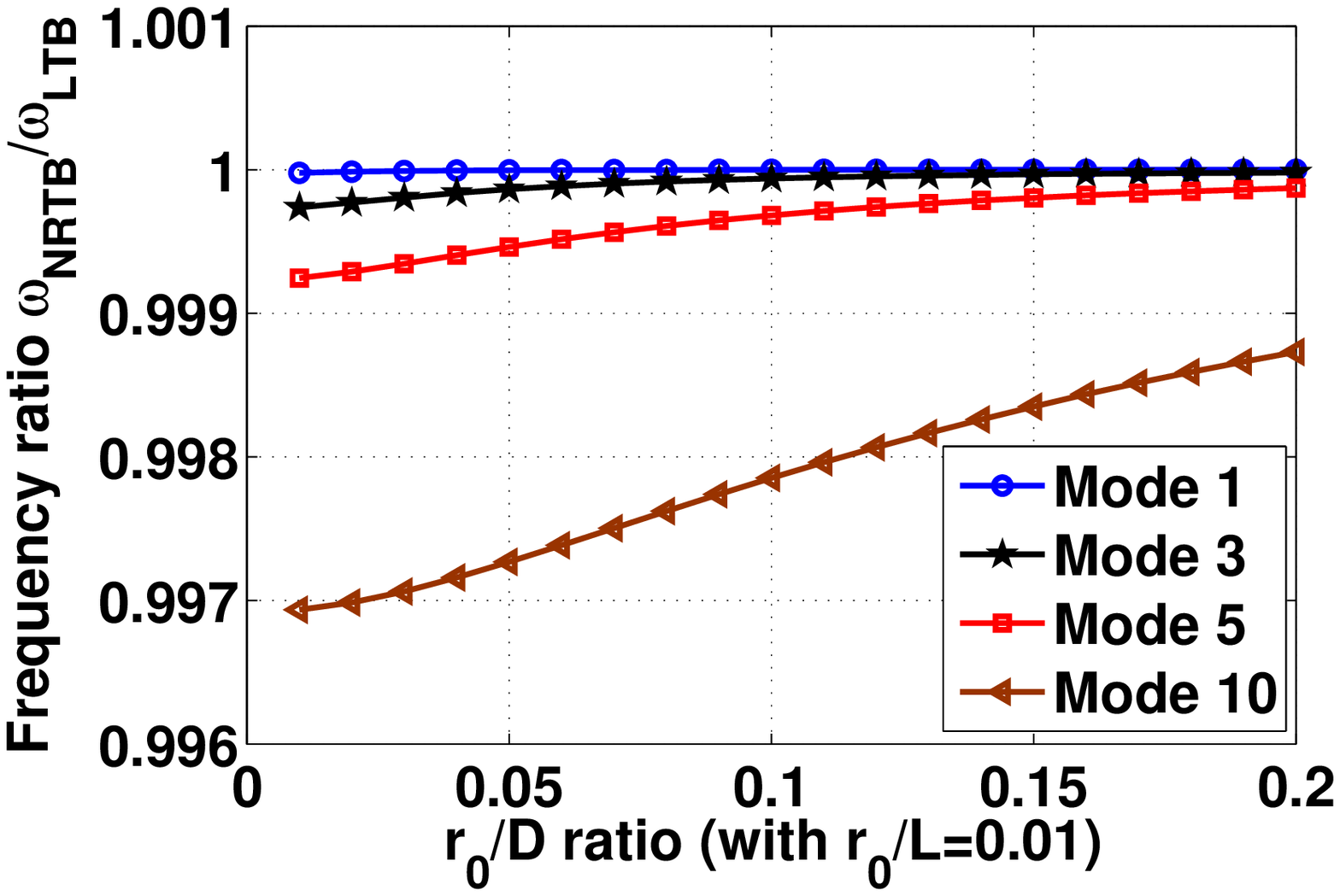}
        }%
        \subfigure[]{%
            \label{ssb4}
            \includegraphics[width=0.45\textwidth]{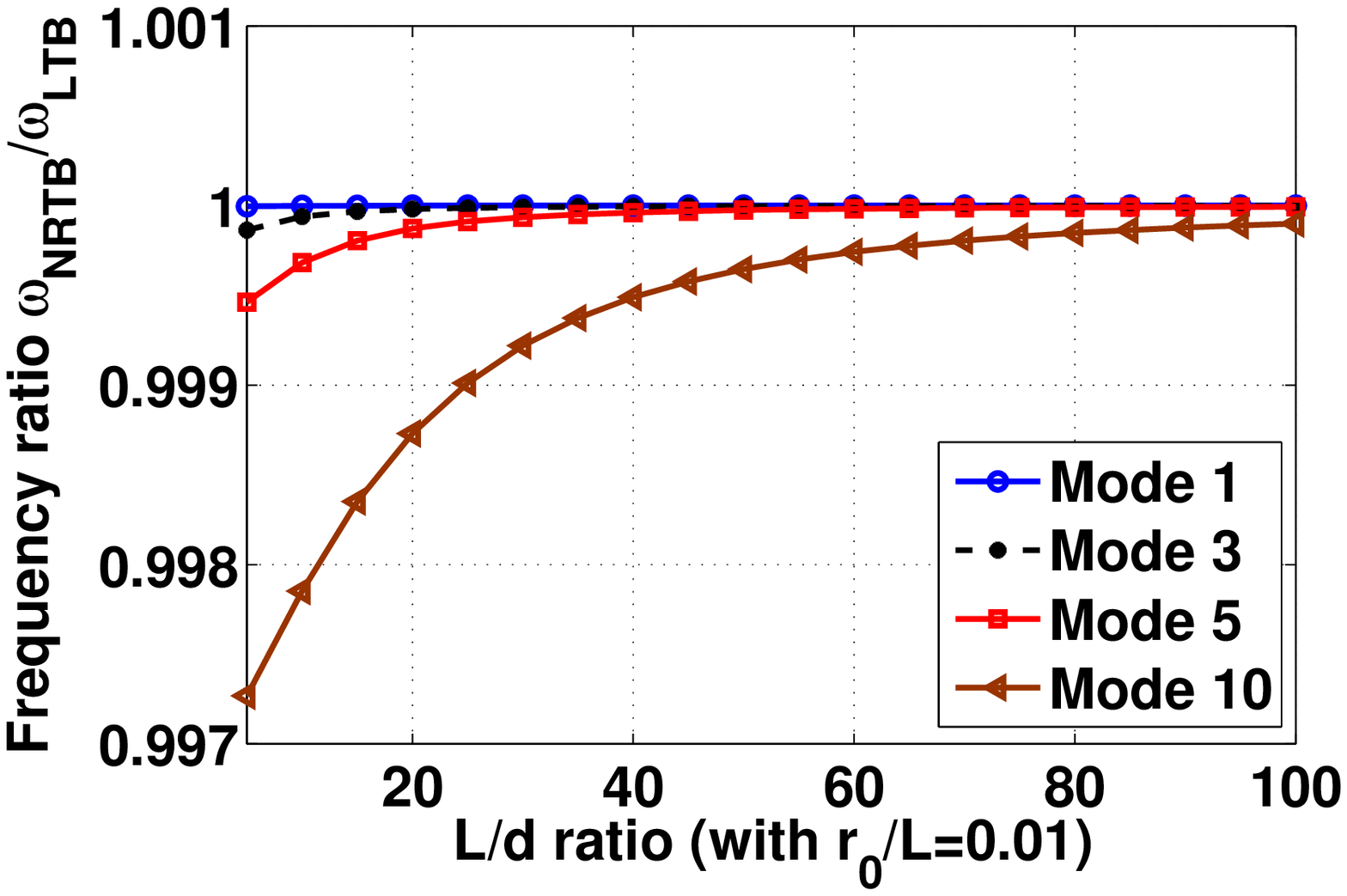}
        }%
    \end{center}
    \caption{%
        Small scale effects on the natural frequency ratios of the simply supported nonlocal rational Timoshenko beam:
        $(a)$ effect of $r_0/L$ ratio on the natural frequency ratios with $r_0/d=0.02$;
        $(b)$ effect of $r_0/L$ ratio on the natural frequency ratios with $L/d=10$;
        $(c)$ effect of $r_0/d$ ratio on the natural frequency ratios with $r_0/L=0.01$;
        and $(d)$ variation of the natural frequency ratios with different values of $L/d$.
        Here, the subscripts NRTB and LTB denote the corresponding values obtained from
        the nonlocal rational Timoshenko beam model and the classical Timoshenko beam model, respectively.
        }%
     \label{RETB2}
\end{figure}

\subsection{Small scale effect on the natural frequencies of the nonlocal rational Timoshenko beam model}
To study the effect of small scale parameter (i.e., the length scale parameter) on the natural frequencies of the nonlocal rational Timoshenko beam (NRTB) model, we consider a simply supported beam with the same material and elastic parameters. A 2D atomistic system of length $L$ and depth $d$ is considered as a simply supported beam. The boundary conditions of simply supported beams are
\begin{align}\label{RETB0b}
v_0(x,t) = 0 \;\;\;\; \mbox{and} \;\;\;\; M(x,t) = EI \Delta_1^2 \frac{\partial \phi_0}{\partial x} = 0  \;\;\;\; \mbox{at} \;\;\;\; x=0,L
\end{align}
To satisfy the simply supported boundary conditions, the solutions of the nonlocal rational Timoshenko beam (Eq. (\ref{SEBE5044b})) are assumed of the form
\begin{align}\label{RETB1b}
v_0(x,t) = \sum_{n=1}^{\infty}\hat{v}_{0n} \sin \frac{n\pi x}{L} e^{i_m \omega_n t}, \;\;\;\; \phi_0(x,t) = \sum_{n=1}^{\infty}\hat{\phi}_{0n} \cos \frac{n\pi x}{L} e^{i_m \omega_n t}
\end{align}
Substituting these solutions into Eq. (\ref{SEBE5044b}) gives
\begin{equation}\label{RETB2b}
  \left[
  \begin{array}{cc}
  GAK_1 S_n^2 - \omega_n^2 \rho A & - GAK_1 S_n \\
  - GAK_1S_n & EI S_n^2 + GAK_1 - \omega_n^2 \rho IK_2\\
  \end{array}
  \right]\left(
         \begin{array}{c}
         \hat{v}_{0n} \\
         \hat{\phi}_{0n}\\
         \end{array}\right)
         =\left(
         \begin{array}{c}
         0 \\
         0 \\
         \end{array}\right)
\end{equation}
where $S_n = (\sin  \frac{n\pi r_x}{2L}/\frac{r_x}{2})$. Setting the determinant of the above matrix to zero, one can obtain the characteristic equation in the general case as
\begin{align}\label{RETB3b}
\rho^2 A IK_2\omega_n^4 - ((GAK_1 \rho I K_2 + EI \rho A)S_n^2 + \rho GA^2K_1)\omega_n^2 + (GAK_1EI) S_n^4= 0
\end{align}
Solving which, one can obtain the expression of real valued natural frequencies as
\begin{align}\label{RETB4b}
\omega_n(n,r_x,L,d)_{1,2} = + \left[ \frac{D_2}{2D_1} \pm \sqrt{\frac{{D_2}^2}{4{D_1}^2} - \frac{D_3}{D_1}}\right]^{\frac{1}{2}}
\end{align}
where $D_1=\rho^2 A IK_2$, $D_2=((GAK_1 \rho I K_2 + EI \rho A)S_n^2 + \rho GA^2K_1)$, and $D_3=GAK_1EI S_n^4$.

The microstructure dependence of natural frequencies of the simply supported beam is studied using Eq. (\ref{RETB4b}). The small scale effects on the natural frequency ratios (i.e., ratio of the frequency of the nonlocal rational Timoshenko beam to the same of local Timoshenko beam) are presented in Fig. \ref{RETB2}. Results show that the natural frequency predictions of NRTB theory are much smaller as compared to the predictions of LTB theory for higher $r_0/L$ ratios with a fixed beam depth (Fig. \ref{ssb1}). Furthermore, NRTB model predicts smaller frequency values with the increasing values of $r_0/L$ ratio for any fixed $L/d$ ratio (Fig. \ref{ssb2}). Fig. \ref{ssb3} shows that the frequency ratios are much lesser than unity for lower $r_0/d$ values with a fixed beam length. It is clearly observed that the NRTB theory predicts smaller natural frequencies for stubbier nano-beams (Figs. \ref{ssb1},\ref{ssb2} and \ref{ssb3}). This is physically very significant with respect to the experimental observations reported in literatures \citep{abbasion2009free,wang2007vibration}. Moreover, the higher mode natural frequencies of NRTB are more sensitive to the variation of $r_0/L$ ratios (Figs. \ref{ssb1},\ref{ssb2}). The variations of the natural frequency ratios with the variation of $L/d$ ratio keeping the $r_0/L$ ratio fixed is presented in Fig. \ref{ssb4}. Therefore, Fig. \ref{ssb4} demonstrates that the effects of transverse shear deformation and rotary inertia on vibration frequency ratios of a simply supported beam. It can be seen in Fig \ref{ssb4} that the frequency ratios are much smaller than unity for all modes at lower $L/d$ ratios. Figs. \ref{ssb3}, \ref{ssb4} show that the predictions of some lower natural frequencies by both the local TB and nonlocal rational Timoshenko beam models converge for very slender beams. These theoretical observations (in good agreement with \cite{abbasion2009free,wang2007vibration}) are physically very significant for dynamics of many realistic nano-electro-mechanical systems. All the above results show interesting capabilities of the unified nonlocal rational continuum models.

\section{Summary}
A contributing development of unified nonlocal rational continuum equations to elucidate the basic mechanics of many heterogeneous atomistic systems is presented in this paper. These nonlocal rational continuum models have only one microstructure-dependent length scale parameter, which can be obtained from experimental observations. The popular strain gradient elasticity models \citep{toupin1962elastic,mindlin1964micro} and integral type nonlocal elasticity models \citep{eringen1983differential} are shown to be special approximations of the unified nonlocal rational continuum model. The other motivating features of the proposed nonlocal rational continuum enrichment technique are as follows:
\begin{itemize}
    \item A novel transformation technique of a discrete differential expression into an exact equivalent rational continuum derivative form is developed on the basis of Taylor's series transformation of the continuous field variables and considering Bloch wave type solutions.
    \item The exact equivalent nonlocal rational continuum model of a 1D harmonic lattice with nearest and/or non-nearest neighbor interactions is developed using the novel discrete to continuum transformation technique.
    \item The corrected form of the ``bad Boussinesq problem" is developed directly from the atomistic equation of a 1D harmonic lattice without any approximations. This corrected equation overcomes the instability and non-convexity problems which are commonly known as the consequences of the sign paradox associated with this ``bad Boussinesq problem".
    \item In the development process of the novel discrete to continuum conversion technique, the discrete dispersion differential operators $\Delta_m$s are derived for different orders of derivatives. Many higher order continuum equations can be enriched very efficiently using these discrete dispersion differential operators $\Delta_m$s. nonlocal rational Mindlin-Herrmann rod and nonlocal rational Timoshenko beam equations are developed as the equivalent higher-order 1D realization of actual 2D atomistic systems.
    \item Analytical solutions for dispersion of 1D nonlocal rational rod equations exactly match with the atomistic solutions. Moreover, dispersion predictions of nonlocal rational Mindlin-Herrmann rod, and nonlocal rational Timoshenko beam equations show good agreement with the actual atomistic solutions.
    \item The nonlocal rational rod equations are solved semi-analytically using NLSFEM (\cite{patra2014spectral}) for displacement and velocity responses which show good agreement with the MD solutions of the actual atomistic systems.
\end{itemize}

The salient features of this elegant continuum enrichment technique are simplicity, generality, and solvability. The above developments substantiate that the peculiarity in dispersive characteristics of any material system is mainly due to the existent heterogeneity and nonlocal interactions between the non-nearest neighbor atoms. These nonlocal rational continuum equations with enhanced dispersive characteristics can be very efficiently used in designing several nano-sensors and devices, phononic metamaterials, and in multiscale modeling of many realistic novel material systems.

\section*{References}
\bibliography{myJMPSbib}

\end{document}